\documentclass[a4paper,11pt]{article}
\usepackage{physics}
\usepackage{jcappub}
\usepackage{pgfplots}

\usepackage{amsmath}
\usepackage{bm}
\usepackage{xcolor}
\usepackage{float}
\usepackage{multirow}
\usepackage{makecell}

\newcommand{\aap}{{Astron.~Astrophys.}}
\newcommand{\apjl}{{Astrophys.~J.~Lett.}}
\newcommand{\apjs}{{Astrophys.~J.~Supp.}}

\usepackage{graphicx}
\usepackage{subcaption}
\usepackage{amssymb, amsmath}
\usepackage[amssymb]{SIunits}
\usepackage{epstopdf}
\usepackage{hyperref}
\usepackage{aas_macros}
\usepackage{xfrac}

\usepackage[mathscr]{eucal}	
\DeclareTextSymbol{\degre}{T1}{23}

\newcommand{\beq} {\begin{equation}}
\newcommand{\eeq} {\end{equation}}
\newcommand{\bal} {\begin{aligned}}
\newcommand{\eal} {\end{aligned}}

\newcommand{\refeq}[1]{Eq.~(\ref{eq:#1})}          
\newcommand{\refeqs}[2]{Eqs.~(\ref{eq:#1})--(\ref{eq:#2})}          
\newcommand{\reffig}[1]{Figure~\ref{fig:#1}}          
\newcommand{\refsec}[1]{Section~\ref{sec:#1}}

\newcommand{\reftab}[1]{Table~\ref{tab:#1}}

\newcommand{\ie}{\emph{i.e.}~}

\newcommand{\etc}{\emph{etc.}}

\title{Testing charge quantization with axion string-induced cosmic birefringence}

\author[a]{Weichen Winston Yin,}
\author[a]{Liang Dai,}
\author[b, a]{and Simone Ferraro}
\affiliation[a]{Department of Physics, 366 Physics North MC 7300, University of California, Berkeley, CA 94720, USA}
\affiliation[b]{Lawrence Berkeley National Laboratory, One Cyclotron Road, Berkeley, CA 94720, USA}
\emailAdd{winstonyin@berkeley.edu}
\emailAdd{liangdai@berkeley.edu}
\emailAdd{sferraro@lbl.gov}

\abstract{
We demonstrate that the Peccei-Quinn-electromagnetic anomaly coefficient $\mathcal A$ can be directly measured from axion string-induced cosmic birefringence by applying scattering transform to the anisotropic polarization rotation of the cosmic microwave background. This breaks the degeneracy between $\mathcal A$ and the effective number of string loops in traditional inference analyses that are solely based on the spatial power spectrum of polarization rotation. Carrying out likelihood-based parameter inference on mock rotation realizations generated according to phenomenological string network models, we show that scattering transform is able to extract enough non-Gaussian information to clearly distinguish a number of discrete $\mathcal A$ values, for instance $\mathcal{A}=\sfrac{1}{9},\,\sfrac{1}{3},\,\sfrac{2}{3}$, in the ideal case of noise-free rotation reconstruction, and, to a lesser but interesting degree at reconstruction noise levels comparable to that expected for the proposed CMB-HD concept. In the event of a statistical detection of cosmic birefringence by Stage III or IV CMB experiments, our technique can be applied to test the stringy nature of the birefringence pattern and extract fundamental information about the smallest unit of charge in theories beyond the Standard Model.
}

\begin{document}

\maketitle


\section{Introduction}\label{sec:introduction}

It has been a fundamental quest in physics to understand the smallest unit of electric charge, from the measurement of the elementary charge $e$ in the oil drop experiment~\cite{millikan1911isolation} to the search for magnetic monopoles as the source of charge quantization~\cite{Cabrera1982}. It has recently been proposed that measuring cosmic birefringence in the cosmic microwave background (CMB) induced by ultralight axion strings is a promising way to probe charge quantization beyond the Standard Model (SM)~\cite{Agrawal_2020}. In this paper, we build on this idea and demonstrate that non-Gaussian statistical inference using scattering transform coefficients can extract this information from the anisotropic polarization rotation in the CMB in a realistic future experiment.

\subsection{Axions and charge quantization}
Axions are neutral pseudo-scalar fields that generically arise from a spontaneously broken Peccei-Quinn symmetry and are often invoked in beyond-the-SM theories.\footnote{Some authors only refer to the QCD axion as ``axion'', but in this work ``axions'' generally include all axion-like particles.} For example, the QCD axion provides one of the most compelling solutions to the strong CP problem~\cite{PecceiQuinn1,PecceiQuinn2} and is a proposed candidate of the astrophysical dark matter~\cite{PRESKILL1983127,ABBOTT1983133,DINE1983137}. The mixed anomaly between the Peccei-Quinn symmetry and SM electromagnetism gives rise to a Chern-Simons coupling between the QCD axion and the photon as an addition to the SM Lagrangian~\cite{Hooft1980}:
\begin{equation}
    \mathcal L \supset \frac{\mathcal A\,\alpha_{\mathrm{em}}}{4\pi f_a}\,a\,F\,\tilde F,
    \label{eq:chern-simons}
\end{equation}
where $a$ is the axion field, $F$ is the electromagnetic field strength tensor, $f_a$ is the periodicity of the field value such that $a$ is identified with $a+2\pi\,f_a$, $\alpha_{\mathrm{em}}$ is the fine structure constant, and $\mathcal A$ is the Peccei-Quinn-electromagnetic anomaly coefficient.

In particular, the dimensionless anomaly coefficient $\mathcal A$ reveals crucial information about the structure of the theory beyond the SM in the following way. While the periodicity $f_a$ is subject to renormalization, $\mathcal A$ is not, so its value is fixed on all energy scales~\cite{Hooft1980}. If the high-energy theory introduces any new electrically charged fermions, then
\begin{equation}
    \mathcal A = \sum_f\,Q_{a,f}\,Q_f^2,
\end{equation}
where $f$  ranges over the new fermions, $Q_{a,f}$ are the corresponding Peccei-Quinn charges, and $Q_f$ are the corresponding electric charges. Since $Q_{a,f}$ are necessarily integers, $\mathcal A$ is an integer multiple of the square of the smallest charge in the theory~\cite{Agrawal_2020}. For any SM-like theory with fractional charges $\sfrac{1}{3}$ and $\sfrac{2}{3}$, such as minimal Grand Unified Theories, $\mathcal A$ is expected to be a multiple of $\sfrac{1}{9}$. See Ref.~\cite{Agrawal_2022} for a detailed discussion on the quantized nature of $\mathcal A$. If $\mathcal A$ is experimentally measured to be close to a ratio of small integers, the possible charge assignments in the beyond-SM theory will be strongly constrained.

Motivated by this, Ref.~\cite{Agrawal_2020} proposes a method to directly measure $\mathcal A$ independently of $f_a$. The axion-photon coupling in \refeq{chern-simons} implies that, as a distant photon travels through the axion field, its polarization rotates by an angle
\begin{equation}
    \Delta\Phi = \frac{\mathcal A\,\alpha_{\mathrm{em}}}{2\pi f_a}\,\Delta a,
\end{equation}
where $\Delta a$ is the net change in axion field value between the end points of the photon's path~\cite{HarariSikivie_1992,CarrollFieldJackiw_1990,Carroll_1998}. This effect, called cosmic birefringence, induces parity violation in the observed CMB polarization anisotropies as CMB photons propagate through the axion-filled cosmic medium~\cite{Lue_1999}. In ordinary situations, $a \ll 2\pi f_a$ as the axion settles down toward the minimum of the potential, rendering birefringence an extremely weak effect. Remarkably, the periodicity of $a$ allows field configurations with topological defects, called cosmic strings, in which $\Delta a \approx n\,2\pi f_a$ for $n\in\mathbb Z$ along a typical line of sight. Each string has a cosmological-scale length, but only has a microscopic size in the transverse directions. The polarization rotation along a typical direction is then greatly enhanced, $\Delta\Phi \approx n\,\mathcal A\,\alpha_{\mathrm{em}}$, which is quantized in a way that depends only on $\mathcal{A}$ but not on $f_a$~\cite{Agrawal_2020}.

Axion field configurations with cosmic strings can naturally arise in our Universe. If the Peccei-Quinn symmetry is broken after cosmic inflation, the axion field value in different causal patches will take unrelated values. Along an imagined superhorizon closed path, the axion field value often changes by one or more periods. Thus, the path has a non-zero axion winding number and must enclose cosmic strings, along which the axion field is singular and the Peccei-Quinn symmetry is restored~\cite{Kibble:1976sj,Kibble:1980mv,Hindmarsh_1995}. Since axion strings are topologically stable, the cosmic string network could have survived until long after recombination.

The primary subject of this work are ultralight axions whose masses are less than the Hubble scale at recombination, $m_a \lesssim 3H_{\mathrm{cmb}} \simeq 8\times 10^{-29}\,\mathrm{eV}$. While such ultralight axions neither solve the strong CP problem nor contribute a significant fraction of the dark matter~\cite{Hu_2000}, they may drive the current accelerated expansion of the universe~\cite{Marsh_2016} and are commonly predicted in string theory scenarios~\cite{Cicoli_2012,Halverson_2019,Broeckel_2021}. For $m_a \gtrsim 3H_0 \simeq 4\times 10^{-33}\,\mathrm{eV}$, the formation of domain walls becomes relevant to the dynamics of the axion string network. For this mass range, we only consider the case in which each string is attached to at least two domain walls ($N_{\mathrm{DW}} > 1$), where the balance of forces ensures the stability of the string-wall network. String-wall networks with $N_{\mathrm{DW}} = 1$ in this mass range collapse and dissipate into a bath of axion radiation at a time before the present day depending on $m_a$ \cite{Chang_1998,Hiramatsu_2013}. The string-wall collapse scenario can in principle be studied with the technique expounded in this work, but we omit it for clarity and simplicity. See Ref.~\cite{Jain_2022} for a detailed study of cosmic birefringence due to axion string-wall networks.

\subsection{CMB polarization rotation}
In a universe teeming with axion cosmic strings, the polarization of each CMB photon rotates by $\pm\mathcal A\,\alpha_{\mathrm{em}}$ if the photon passes through a string loop or by $\pm\frac 12\mathcal A\alpha_{\mathrm{em}}$ if it passes by a long open string~\cite{Agrawal_2020}. The stacked effect of the entire string network will manifest in the CMB as the division of the sky into many domains of nearly uniform polarization rotation. The rotation angle in each domain will be quantized as integer multiples of $\mathcal A\,\alpha_{\mathrm{em}}$.

If the anistropic polarization rotation of the CMB could be measured with arbitrary precision at arbitrarily high spatial resolution, then any individual cosmic string would be resolved. The value of $\mathcal A$ can be extracted from the difference in the polarization rotation angle on both sides of the string. However, since the intrinsic CMB polarization along any line of sight is not known, the rotation angle has to be statistically inferred from the correlation between Fourier modes on the sky. Current experiments (\emph{e.g.}~Planck) do not measure sufficiently many signal-dominated Fourier modes to locally detect a single string for interesting values of $\mathcal{A}$. Instead, statistical detection based on accumulated significance over a large sky patch containing multiple strings is a more promising approach~\cite{Agrawal_2020}.

Motivated by numerical simulations, Ref.~\cite{Jain_2021} developed a phenomenological model of cosmic string network called the loop-crossing model, which can be used to efficiently generate random realizations of birefringence and semi-analytically calculate its power spectrum. On the other hand, previous studies have employed quadratic estimators for statistical detection and measurement of the polarization rotation, and the power spectrum of these quadratic estimators is well understood~\cite{Yadav_2009,Yin_2022}.

Besides $\mathcal A$, the loop-crossing model introduces other phenomenological parameters that characterize the cosmic string network, including $\xi_0$, the effective length of strings per Hubble volume in Hubble units. In a previous work~\cite{Yin_2022}, we applied quadratic estimators to the loop-crossing model and forecast detectability of a string network by forthcoming CMB experiments. Quoting binned rotation power spectrum derived from the Planck 2015 polarization data~\cite{planck_constraint_2017}, we constrain $\mathcal A^2\,\xi_0 < 0.93$ at a 95\% confidence level~\cite{Yin_2022}. Upcoming experiments (Simons Observatory, CMB-S4, \etc) will be sensitive enough to discover or falsify anisotropic rotations from an axion string network in the theoretically plausible parameter space $\mathcal{A} = 0.1\sim 1$ and $\xi_0 = 1\sim 100$.

\subsection{Breaking the degeneracy between the strength and number of strings}
Analyses based on the rotation power spectrum unfortunately suffer from a shortfall for phenomenological models like the loop-crossing model. Since the power spectrum is optimized for Gaussian random fields but misses all non-Gaussian spatial information in the rotation field, it is only sensitive to $\mathcal A^2\,\xi_0$, a combination of ``strength'' and ``quantity'' of the string network, but otherwise is unable to constrain $\mathcal A$ or $\xi_0$ separately. Anticipating a future detection of cosmic birefringence from axion strings, and given the theoretical significance of $\mathcal A$, it will be highly rewarding to develop a methodology that breaks the degeneracy between $\mathcal{A}$ and $\xi_0$, which will confirm the string origin of the anisotropic rotation and place tight constraints on $\mathcal{A}$.

We suggest in this paper alternative summary statistics that extract a substantial amount of non-Gaussian information. Scattering transform, which manipulates input fields in a way similar to what a convolutional neural network does but requires no training, offers precisely this advantage over the power spectrum. In recent applications to cosmology, by exploiting non-Gaussian features in the weak lensing field~\cite{Cheng_2020, Cheng_2021} or in the large-scale structure~\cite{Valogiannis_2022}, scattering transform is shown to significantly reduce the degeneracy between cosmological parameters at the power spectrum level, such as $\sigma_8$ and $\Omega_m$.

We show in this paper that in the event of a detection of axion string birefringence at Stage III and/or Stage IV CMB experiments using quadratic estimators, scattering transform can be further employed to break the degeneracy between $\mathcal A$ and $\xi_0$, 
for noise levels achievable by the conceived next-generation high-sensitivity experiment CMB-HD~\cite{Ferraro_2022}. The methodology may distinguish different discrete values of $\mathcal A$, which will provide a strong test of charge quantization beyond the SM.

\subsection{Outline}
The remainder of this paper is organized as follows. In \refsec{models}, we briefly review the loop-crossing model, explains how it is realized on the flat sky, and shows that the $\mathcal A^2\,\xi_0$ degeneracy arises in the power spectrum. In \refsec{scattering}, we discuss the technique of scattering transform and compare it with the power spectrum analysis, guided by understanding of the underlying mathematics. In \refsec{noise}, we design a numerical study in which the mock reconstruction noise is generated and added to the string birefringence signal. In \refsec{method}, we detail the parameter inference procedure. In \refsec{results}, we forecast inference results for both noise-free and noisy rotation field, with a comparison between scattering transform and power spectrum analysis. Throughout this work, we adopt $c=1$ units.

\section{Loop-crossing model}
\label{sec:models}

The spatial pattern of string induced birefringence in the CMB depends on the string network structure which is dictated by string dynamics. While the precise distribution of string loops as a function of loop size and redshift is the subject of ongoing numerical investigations~\cite{Blanco_Pillado_2014,Gorghetto_2018,Buschmann_2022}, phenomenological string network models are useful to approximate the range of cosmic string networks found by physical simulations, and they enable efficient comparison between mock data and theoretical predictions in forecast studies.

One example is the loop-crossing model developed in Ref.~\cite{Jain_2021} , in which the Universe is populated by circular string loops with a redshift dependent radius distribution. At a given redshift $z$, the comoving number density of string loops of comoving radii in the interval $[r,\,r+{\rm d}r]$ is
\begin{equation}
    \mathrm dn = \nu(r, z)\,\mathrm dr.
\end{equation}
String loop orientations are assumed to be random and isotropic. It is convenient to re-parametrize the radius distribution $\nu(r,\,z)$ as
\begin{equation}
    \nu(r, z) = \frac{[a(z)\,H(z)]^4}{2\pi\,\zeta}\chi(\zeta,\,z),
\label{eq:distribution_reparam}
\end{equation}
where the dimensionless parameter $\zeta = r\,a\,H$ is the proper radius of a string loop expressed in units of the Hubble length at redshift $z$. The radius distribution $\chi(\zeta, z)$ completely characterizes the string network in the loop-crossing model.

In what follows, we will consider two different loop radius distributions in which the energy density in the cosmic string network scales with the dominant energy density in the Universe. This stable distribution is achieved through the dynamical process of string motion and recombination (not to be confused with cosmological recombination)~\cite{Gorghetto_2018}. The resultant network is said to be ``in scaling", and its distribution of dimensionless string radii is redshift-independent, $\chi(\zeta,\,z) = \chi(\zeta)$. Although this property of axion string networks has recently been called into question~\cite{Gorghetto_2018,Buschmann_2022}, the method described in this paper would only need minimal modification by simply allowing for a redshift-dependent string radius distribution $\chi$.

The first model (Model I hereafter) assumes that all string loops have the same radius in Hubble units $\zeta = \zeta_0$:
\begin{equation}
    \chi_{\mathrm I}(\zeta) = \xi_0\,\delta(\zeta-\zeta_0).
    \label{eq:model1_distrib}
\end{equation}
The second model (Model II hereafter) assumes that a fraction $f_{\mathrm{sub}}$ of the loops are sub-Hubble scale loops, distributed logarithmically between a variable $\zeta_{\mathrm{min}}$ and $\zeta_{\mathrm{max}}=1$, while the rest of the loops have a Hubble-length radius:
\begin{equation}
    \chi_{\mathrm{II}}(\zeta) = (1-f_{\mathrm{sub}})\,\xi_0\, \delta(\zeta-\zeta_{\mathrm{max}}) + f_{\mathrm{sub}}\, \xi_0\frac{\Theta(\zeta_{\mathrm{max}}-\zeta)\,\Theta(\zeta-\zeta_{\mathrm{min}})}{\zeta_{\mathrm{max}}-\zeta_{\mathrm{min}}}.
    \label{eq:model2_distrib}
\end{equation}
Model I and Model II become the same if $f_{\mathrm{sub}}=0$ and $\zeta_0 = \zeta_{\mathrm{max}} = 1$. The logarithmic loop length distribution of Model II is consistent with recent simulations~\cite{Buschmann_2022}.

Both distributions are normalized to
\begin{equation}
    \int_0^\infty\,\chi(\zeta)\,{\rm d}\zeta = \xi_0.
\end{equation}
The parameter $\xi_0$ can be interpreted as the number of Hubble-scale loops per Hubble volume there would have to be in order to account for the same mean energy density as in the string network described by the distribution $\chi(\zeta)$. It is therefore called the effective number of strings per Hubble volume.

In the parameter inference procedure of \refsec{method}, we will treat $(\mathcal A,\,\xi_0,\,\zeta_0)$ as the parameter space of Model I and $(\mathcal A,\,\xi_0,\,f_{\mathrm{sub}})$ as the parameter space of Model II, and set $\zeta_{\mathrm{min}} = 10^{-1}$. We note that a smaller $\zeta_{\mathrm{min}}$ would lead to a reduction of the birefringence signal.

\subsection{Implementation of models}
\label{sec:realization}

We describe the procedure for generating random realizations of polarization rotation in the CMB generated by a string network described by a given loop radius distribution $\chi(\zeta,\,z)$, where the $z$-dependence is kept for generality. The procedure consists of three steps:
\begin{enumerate}
    \item Determine the average number of string loops whose centers lie within a chosen patch on the sky.
    \item Sample the redshifts and radii of string loops according to the distributions specified by the model, and draw loop centers uniformly within the chosen patch of the sky.
    \item Calculate the rotation pattern due to each individual string loop. Find the superposition of constributions from all generated string loops.
\end{enumerate}
We work in the flat-sky approximation for simplicity. A similar process taking into account the curvature of the sky can be found in Ref.~\cite{Jain_2022}.

Using the parametrization in \refeq{distribution_reparam}, the comoving number density of string loops is
\begin{equation}
    \mathrm d n = \frac{(a\,H)^3}{2\pi\,\zeta}\,\chi\,{\rm d}\zeta.
\end{equation}
The specific number of string loops in the redshift interval between $z$ and $z+\mathrm dz$ and within a solid angle $\mathrm d\Omega$ on the sky is
\begin{equation}
    \mathrm dn\,\mathrm dV_C = \frac{(a\,H)^3\,s^2}{2\pi\,\zeta\,H}\,\chi\,\mathrm d\zeta\,\mathrm dz\,\mathrm d\Omega,
\label{eq:number_in_volume}
\end{equation}
where $\mathrm dV_C$ is the comoving volume within solid angle $\mathrm d\Omega$ in the redshift interval between $z$ and $z+\mathrm dz$, and $s$ is the comoving distance out to redshift $z$. The expected number of string loops observed within a patch of the sky of solid angle $\Omega$ is obtained by integrating over radius, redshift, and solid angle:
\begin{equation}
    \langle n\rangle = \Omega\int_0^1 \mathrm d\zeta \int_0^{z_{\mathrm{cmb}}} \mathrm dz\, \frac{(a\,H)^3\,s^2}{2\pi\,\zeta\,H}\chi,
\label{eq:poisson_rate}
\end{equation}
where we assume no string loop exceeds the Hubble scale, \ie $0<\zeta\le 1$. In practice, the actual number of string loop centers within a finite patch of sky is drawn from a Poisson distribution with a mean number $\langle n\rangle$.

We may simplify the expression when $\chi(\zeta,\,z)$ has a separable functional form:
\begin{equation}
    \chi(\zeta,\,z) = X(\zeta)\,Y(z).
\end{equation}
Define rescaled variables $\tilde\zeta = \tilde\zeta(\zeta)$ and $\tilde z = \tilde z(z)$ such that
\begin{equation}
    \mathrm d\tilde\zeta = \frac{1}{2\pi\,\zeta}\,X(\zeta)\,\mathrm d\zeta,\quad \mathrm d\tilde z = \frac{(a\,H)^3\,s^2}{H}\,Y(z)\,\mathrm dz.
\end{equation}
Then, \refeq{number_in_volume} becomes
\begin{equation}
    \mathrm dn\,\mathrm dV_C = \mathrm d\tilde\zeta\,\mathrm d\tilde z\,\mathrm d\Omega,
\label{eq:uniform_distrib}
\end{equation}
and \refeq{poisson_rate} becomes
\begin{equation}
    \langle n\rangle = \Omega\,\Delta\tilde\zeta\,\Delta\tilde z,
\end{equation}
where $\Delta\tilde\zeta$ and $\Delta\tilde z$ are the ranges of the transformed variables corresponding to the ranges of $\zeta$ and $z$ in \refeq{poisson_rate}.

With \refeq{uniform_distrib}, the radii and redshifts of the string loops can be generated by first drawing uniformly in the $(\tilde\zeta,\,\tilde z)$-space within the ranges $\Delta\tilde\zeta$ and $\Delta\tilde z$, and then converting the values back to the $(\zeta,\,z)$-space.

For each string loop with parameters $(\zeta,\,z)$, we calculate its induced spatial rotation pattern in the sky. Since $\zeta/a\,H$ is the proper radius of the loop, its angular radius is $\tan^{-1}(\zeta/a\,H\,s)$~\footnote{We do not approximate $\tan^{-1} x \sim x$ due to the large size of string loops at low redshift.}. The string loop center is uniformly drawn from within the solid angle $\Omega$ of the sky. The unit vector $\hat{\mathbf k}$ normal to the plane of the string loop circle, which is sampled from an isotropic distribution, defines an apparent ellipse as the projection of this circle along the line of sight. Finally, the rotation field is assigned the value of 0 outside the ellipse and $\pm\mathcal A\,\alpha_{\mathrm{em}}$ inside, where the sign depends on whether $\hat{\mathbf k}$ points away from or towards the observer. The total birefringence signal of a string network is simply the sum of signals due to individual string loops.

In practice, we draw the coordinates of string loop centers from a larger sky area $\Omega' > \Omega$ than the one analyzed, so as to include string loops that overlap with $\Omega$ but have centers outside of $\Omega$. To determine $\Omega'$, we set a lower limit on the string redshift $z_{\mathrm{min}} = 0.001$ and calculate an upper limit on the angular radius.

\begin{figure}
    \begin{center}
        \scalebox{0.65}{\input{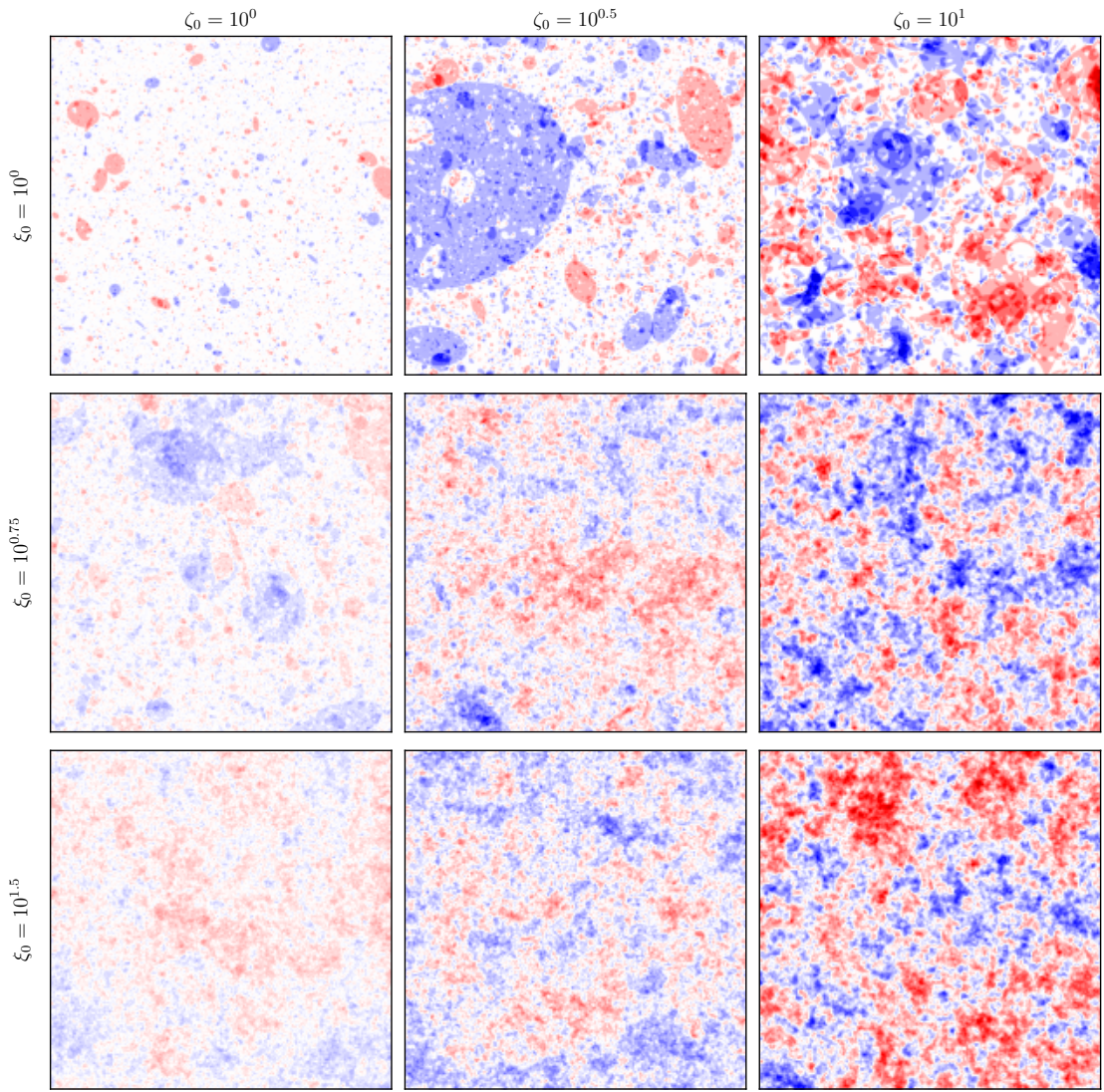}}
    \end{center}
    \caption{Uncurated noise-free realizations of Model I (uniform radius). We plot the rotation field in a $128^\circ\times 128^\circ$ patch of sky for a selection of model parameters, keeping $\mathcal A^2\,\xi_0=1$. With all other parameters fixed, the rotation field scales linearly with $\mathcal{A}$.}
    \label{fig:realization_model1}
\end{figure}

We show in \reffig{realization_model1} several uncurated noise-free realizations of Model I (uniform radii). We plot the rotation field in a $128^\circ\times 128^\circ$ patch of sky (40\% of the full sky) for a selection of model parameter sets, while fixing the overall scaling $\mathcal A=1$. Note that if we rescale each map in \reffig{realization_model1} to keep $\mathcal A^2\,\xi_0$ constant, realizations in the same column (with the same $\zeta_0$) correspond to the same power spectrum. We expect the scattering transform to break this degeneracy. It is visible to the naked eye that the rotation field has more contiguous patches of (nearly) constant rotation angle at lower $\xi_0$ and $\zeta_0$. This is because a low number density of string loops (the expected number of string loops is proportional to $\xi_0$) allows for large patches of coherent rotation, while the smaller radius $\zeta_0$ minimizes the chance that the largest projected string loops intersect with other large string loops. These contiguous features render the rotation field significantly non-Gaussian. We therefore expect that scattering transform analysis can mitigate the parameter degeneracy present in power spectrum analysis, and that the improvement is more pronounced in this part of the parameter space.

\begin{figure}
    \begin{center}
        \scalebox{0.65}{\input{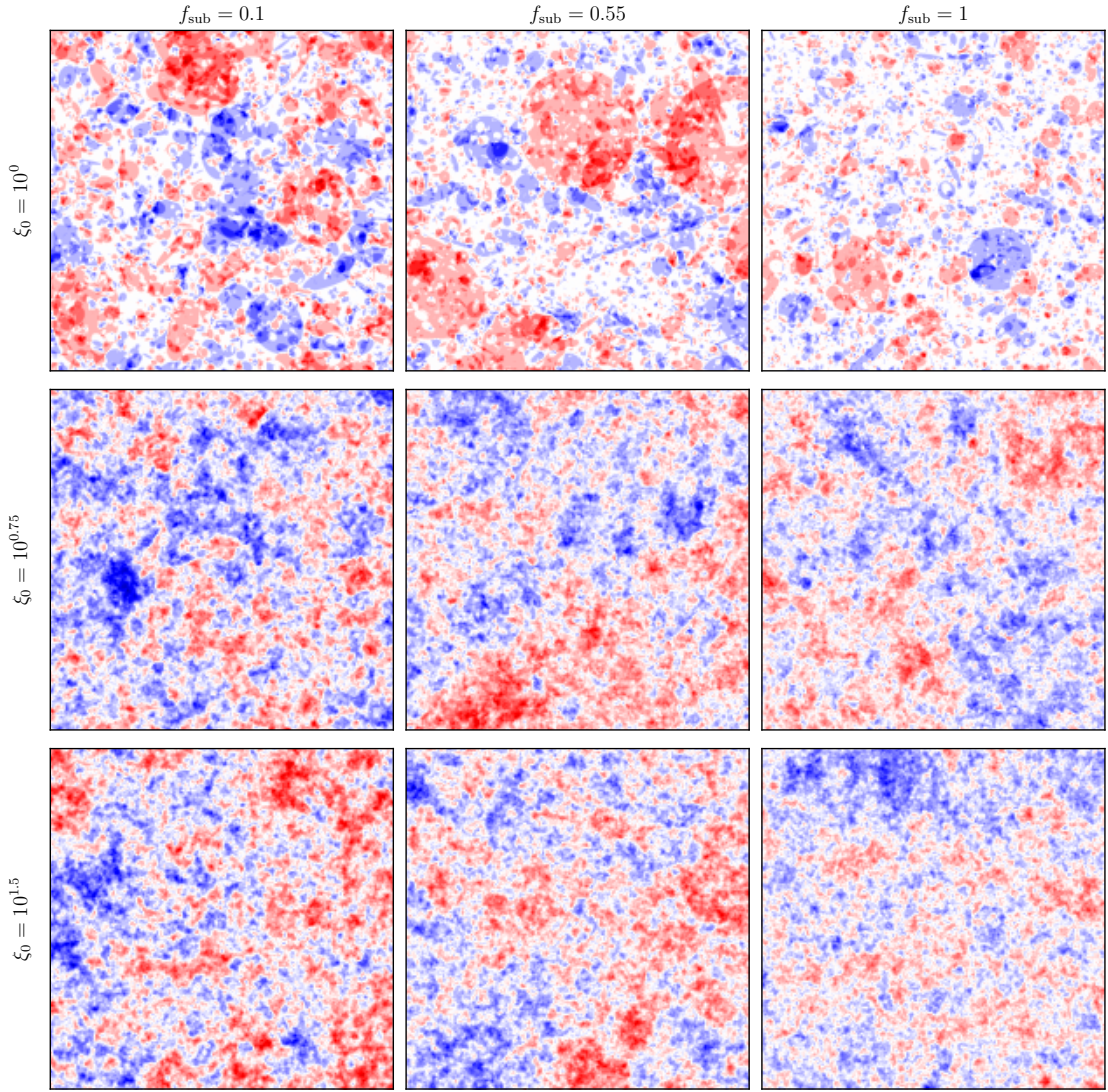}}
    \end{center}
    \caption{Uncurated noise-free realizations of Model II (log-flat radius distribution). We plot the rotation field in a $128^\circ\times 128^\circ$ patch of sky for a selection of model parameters, keeping $\mathcal A^2\,\xi_0=1$.}
    \label{fig:realization_model2}
\end{figure}

Similarly, we show in \reffig{realization_model2} several uncurated realizations of Model II (log-flat radius distribution). Realizations having smaller $\xi_0$ but larger $f_{\mathrm{sub}}$ appear more non-Gaussian. Due to a wide range of string loop radii, the degree of non-Gaussianity in this model is less obvious than Model I with a unique string loop radius.

\subsection{Power spectrum degeneracy}
\label{sec:degeneracy}

Studies on the detection of the CMB birefringence signal have thus far focused on exploiting Gaussian information through the power spectrum analysis~\cite{Yadav_2009,Yin_2022}. Given the Fourier transform of a polarization rotation field $\alpha(\mathbf L)$, the (isotropic) power spectrum of the rotation angle, $C^{\alpha\alpha}_L$, is defined through
\begin{equation}
    \langle\alpha(\mathbf L)\,\alpha^*(\mathbf L')\rangle = (2\pi)^2\,\delta^{(2)}(\mathbf L-\mathbf L')\,C^{\alpha\alpha}_L,
\end{equation}
where $\delta^{(2)}$ denotes the two-dimensional Dirac delta function. It is a general feature of phenomenological string network models that the birefringence power spectrum only depends on the combination $\mathcal A^2\,\xi_0$ but not on $\mathcal A$ or $\xi_0$ separately.

The total polarization rotation field $\alpha$ is the superposition of $n$ independent and identically distributed fields $\alpha_i$, each due to a single cosmic string. It is also reasonable to assume that $\langle\alpha_i\rangle = 0$, as cosmic strings of opposite orientations about the plane of the sky must occur equally likely. For a fixed string loop number $n$:
\begin{equation}
    \langle\alpha(\mathbf L)\,\alpha^*(\mathbf L')\rangle = \left\langle\sum_{i=1}^n\,\sum_{j=1}^n \alpha_i(\mathbf L)\,\alpha^*_j(\mathbf L')\right\rangle = n\,\langle\alpha_1(\mathbf L)\,\alpha_1^*(\mathbf L')\rangle.
\end{equation}
If instead $n$ is drawn from a Poisson distribution with a rate $\langle n\rangle$, then the same logic leads to
\begin{equation}
    \langle\alpha(\mathbf L)\,\alpha^*(\mathbf L')\rangle = \sum_{k=0}^\infty \frac{\langle n\rangle^k e^{-\langle n\rangle}}{k!}\left\langle\sum_{i=1}^k\,\sum_{j=1}^k \alpha_i(\mathbf L)\,\alpha^*_j(\mathbf L')\right\rangle = \langle n\rangle \,\langle\alpha_1(\mathbf L)\,\alpha_1^*(\mathbf L')\rangle.
\end{equation}
The expected string loop number $\langle n\rangle$ is proportional to $\xi_0$, the effective number of horizon-scale loops per Hubble volume; each $\alpha_i$ strictly scales linearly with $\mathcal A$. We therefore conclude that
\begin{equation}
    C^{\alpha\alpha}_L \propto \mathcal A^2\,\xi_0.
\end{equation}
This exact property is true for the loop-crossing model, a fact already pointed out in Ref.~\cite{Jain_2021}.

Since $\mathcal A$ is a fundamental parameter of new physics at high energy scales that informs us about charge assignment while $\xi_0$ depends on the dynamic evolution of the string network over cosmic times, it is important to break the degeneracy between $\mathcal A$ and $\xi_0$. Given the deficiency of any power spectrum-based inference method in this regard, we are motivated to study alternative inference techniques that exploit a substantial amount of non-Gaussian information.

\section{Scattering Transform}
\label{sec:scattering}

In this section, we describe scattering transform and motivate its use as a parameter inference technique that can efficiently exploit non-Gaussian information that is unavailable to power spectrum-based methods. We refer to Ref.~\cite{Cheng_2020} for a more detailed exposition of scattering transform, where this technique is applied to other problems in cosmology.

Like in power spectrum analysis, scattering transform takes in an input field $I_0(\mathbf x)$, the polarization rotation field $\alpha(\mathbf x)$ in our case, and outputs a set of coefficients that serve as summary statistics used for parameter estimation. When computing the power spectrum, the input field is convolved with a family of plane wave mode functions $\phi_{\mathbf k}(\mathbf x) = e^{i\mathbf k\cdot\mathbf x}$ to produce a set of fields:
\begin{equation}
    P_{\mathbf k}(\mathbf x) = \langle|I_0*\phi_{\mathbf k}|^2\rangle(\mathbf x),
    \label{eq:power_spectrum}
\end{equation}
where the average is taken over random realizations of the input field. It is non-zero only when $\mathbf x=0$, thus only $P_{\mathbf k}(0)$ is a meaningful quantity. If statistical isotropy holds for the input fields, one further reduces this set by averaging over the direction of the Fourier wave vector $\mathbf k$ to obtain the coefficients $P(k) = \langle P_{\mathbf k}(0)\rangle_{\hat{\mathbf k}}$.

In scattering transform, the mode functions used for convolution are spatially localized wavelets, rather than plane waves. One common choice are the Morlet wavelets---plane waves modulated by a 2D Gaussian envelope. Given a template wavelet $\psi(\mathbf x)$, a family of wavelets $\psi^{j,l}(\mathbf x)$ are generated by dilation (labelled by $j=1,\cdots,J-1$) and rotation (labelled by $l=1,\cdots,L$). Convolution then leads to a set of fields
\begin{equation}
    I_1^{j,l}(\mathbf x) = \langle |I_0*\psi^{j,l}|\rangle(\mathbf x),
    \label{eq:scattering-transform-first-order}
\end{equation}
where the average is again performed over random map realizations. If statistical homogeneity and isotropy hold for the input fields, one obtains first-order reduced coefficients by averaging over wavelet position and orientation:
\begin{equation}
    s_1^j = \langle I_1^{j,l}\rangle_{\mathbf x,l}.
    \label{eq:ST1}
\end{equation}

As pointed out in Ref.~\cite{Cheng_2020}, the similarity between \refeq{power_spectrum} and \refeq{scattering-transform-first-order} means that the information contained in first-order scattering transform coefficients is similar to that in binned power spectrum. However, whereas the power spectrum has fine-grained information in the Fourier space ($N_{\mathrm{pix}}/2$ coefficients for a square input field $N_{\mathrm{pix}}$ pixels across), scattering transform only gathers course-grained information ($\log_2 N_{\mathrm{pix}}$ reduced coefficients). This means that scattering transform is expected to have a weaker constraining power than binned power spectrum in the direction orthogonal to the $\mathcal A^2\,\xi_0$ degeneracy. As we demonstrate in \refsec{results}, the best parameter inference procedure is one that combines both summary statistics.

The above scattering transform process can be repeated to produce the second-order reduced coefficients by treating first-order fields produced by convolution as input fields:
\begin{equation}
    I_2^{j_1,l_1,j_2,l_2}(\mathbf x) = \langle |I_1^{j_1,l_1}*\psi^{j_2,l_2}|\rangle(\mathbf x),\quad s_2^{j_1,j_2} = \langle I_2^{j_1,l_1,j_2,l_2}\rangle_{\mathbf x,l_1,l_2}.
    \label{eq:ST2}
\end{equation}
In principle, the procedure can be iterated to produce higher-order convoluted fields and the associated high-order reduced coefficients. In this work, scattering transform coefficients up to only the second-order will be used in our analysis.

Power spectrum, as well as higher-order $N$-point moments, have the following disadvantages compared to scattering transform. The power spectrum is insensitive to the statistical deviation of the input field from a Gaussian random field, and in string network models it has a degenerate dependence on $\mathcal A$ and $\xi_0$ (\refsec{degeneracy}). Higher-order $N$-point moments in principle supply the non-Gaussian information, but they suffer from computational complexity and slow convergence for highly non-Gaussian fields (such as the ones resulting from the loop crossing model). The number of coefficients that must be computed increases dramatically with $N$, with useful non-Gaussian information greatly diluted amongst them. In practice, reliable computation of the higher-order moments also suffers from the problem of large sample variance. Scattering transform, on the other hand, is expected to be very sensitive to highly non-Gaussian input fields, with key non-Gaussian information efficiently captured by only a small set of first- and second-order coefficients~\cite{Cheng_2020}.

\reffig{realization_model1} and \reffig{realization_model2} have shown that, for some model parameters, cosmic strings imprint a highly non-Gaussian rotation pattern in the CMB. The non-Gaussian features are primarily the result of large coherent patches corresponding to a handful of string loops of large angular size. Scattering transform should be well suited for breaking the degeneracy between $\mathcal{A}$ and $\xi_0$ by quantifying the degree of non-Gaussianity as $\xi_0$ varies.

\begin{figure}
    \centering
    \includegraphics[width=0.6\textwidth]{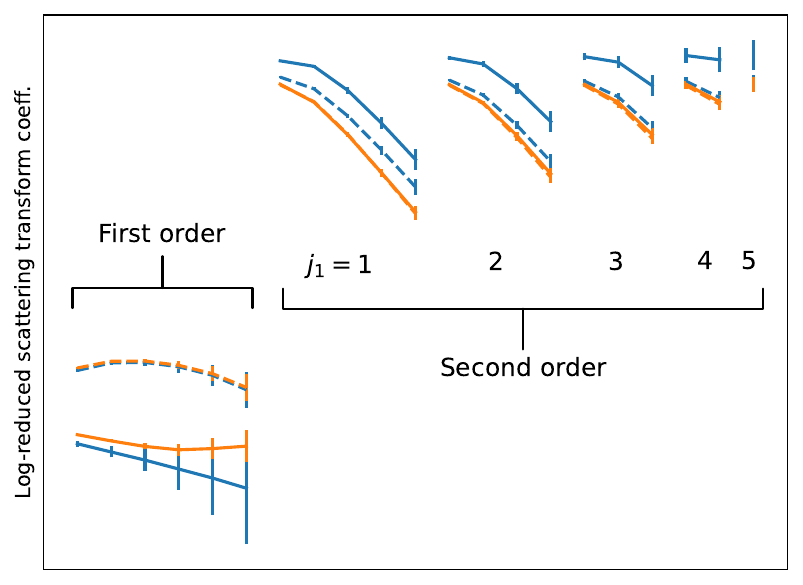}
    \caption{First- and second-order log-reduced scattering transform coefficients, $\log s_1^{j_1}$ and $\log(s_2^{j_1,j_2}/s_1^{j_1})$ across the parameter space of the loop-crossing model. Data points and error bars correspond to the means and standard deviations of coefficients measured from a large number of noise-free realizations. Blue and orange curves correspond to effective string loop numbers of $\xi_0=1$ and $\xi_0=10^{1.5}$, respectively. Solid and dashed curves correspond to string loop radii of $\zeta_0=0.1$ and $\zeta_0=1$, respectively. In all cases, we fix $\mathcal A^2\,\xi_0=1$, which is a degenerate combination in the power spectrum analysis. It is the second-order coefficients that more significantly distinguish between models through their non-Gaussian information.}
    \label{fig:ST_coefs}
\end{figure}

To show how scattering transform sensitively extracts parameters of the loop-crossing model, we plot $\log s_1^{j_1}$ and $\log(s_2^{j_1,j_2}/s_1^{j_1})$ for different parameter values in \reffig{ST_coefs}, assuming the case of a unique string loop radius. Both the mean and standard deviation of each coefficient are plotted, computed from a large set of realizations. By fixing the combination $\mathcal A^2\,\xi_0$ and varying $\xi_0$ and $\zeta_0$ separately, we see that first-order scattering transform coefficients are able to strongly distinguish different $\zeta_0$ but not different $\xi_0$, the same limitation as power spectrum. However, second-order coefficients show great promise at distinguishing different $\xi_0$ with a fixed $\mathcal A^2\,\xi_0$, a marked improvement over the power spectrum analysis.

\section{Reconstruction noise}
\label{sec:noise}

The polarization rotation along any individual line of slight is not directly measurable by instruments. Instead, direction-dependent rotation angles are statistically estimated from the observed CMB primary anisotropies in temperature ($T$) and polarization ($E$ and $B$). This is commonly done via the quadratic estimator (QE), which is an optimally weighted sum of quadratic combinations of the observed $T,E,B$ anisotropy modes to give an unbiased, least-variance estimate of the anisotropy modes of the rotation field~\cite{Yadav_2009}.

Both the conventional power spectrum analysis~\cite{Agrawal_2020, Yin_2022} and the proposed scattering transform method take the \emph{estimated} rotation field as the input field, which includes a noise component resulting from both the statistical variance inherent to the estimator and the instrumental noise of observation. In other words, if $\alpha_o(\mathbf L)$ denotes the quadratic estimation of one Fourier mode of the true polarization rotation field $\alpha$ with a Fourier wave vector $\mathbf L$, then
\begin{equation}
    \alpha_o(\mathbf L) = \alpha(\mathbf L) + \alpha_n(\mathbf L),
\end{equation}
where $\alpha_n(\mathbf L)$ is the noise component. We assume this noise component does not correlate with the true signal $\alpha(\mathbf L)$, and that it is a Gaussian random field entirely determined by a noise power spectrum:
\begin{equation}
    \langle \alpha_n(\mathbf L)\, \alpha^*_n(\mathbf L')\rangle = (2\pi)^2\,\delta^{(2)}(\mathbf L-\mathbf L')\,N(L).
\end{equation}
The noise spectrum $N(L)$ depends on the exact choice of quadratic estimator and instrumental parameters, but it generally dominates over the signal power spectrum $C^{\alpha\alpha}_L$ at large multipoles $L$, where
\begin{equation}
    \langle\alpha(\mathbf L)\,\alpha^*(\mathbf L')\rangle = (2\pi)^2\,\delta^{(2)}(\mathbf L-\mathbf L')\,C^{\alpha\alpha}_L.
\end{equation}
For example, using Model II (see \refsec{models}) with $f_{\mathrm{sub}} = 0.6$ and $\mathcal A^2\,\xi_0 = 1$, the noise spectrum of the Hu-Okamoto estimator dominates over the signal when $L\gtrsim 25$ for Planck SMICA and $L\gtrsim 100$ for CMB-S4~\cite{Yin_2022}.

After producing noise-free realizations of the polarization rotation signal as described in \refsec{realization}, we generate a Gaussian random field $\alpha_n(\mathbf L)$ following a noise spectrum $N(L)$ according to the instrumental parameters\footnote{We choose the $EB$ quadratic estimator for calculating $N(L)$, as it in practice has comparable reconstruction noise level to the Hu-Okamoto estimator and the global minimum variance estimator~\cite{Yin_2022}.}, and then linearly superimpose it onto the rotation signal $\alpha(\mathbf L)$ to obtain a mock observed (estimated) rotation field $\alpha_o(\mathbf L)$.

\section{Method}
\label{sec:method}

The procedure for parameter inference using any summary statistics consists of an observed input field $\alpha_o(\mathbf x)$, the summary statistics $\mathbf d[\alpha_o]$ computed from $\alpha_o$, the covariance matrix $C_{\mathbf d}$ of the summary statistics of $\alpha_o$, and the theoretical summary statistics as a function of model parameters $\mathbf d_{\rm th}(\theta)$. The model parameters for $\alpha_o$ are estimated by maximizing the following log-likelihood function:
\begin{equation}
    \ln\mathcal L[\alpha_o | \theta] = -\frac 12 (\mathbf d[\alpha_o] - \mathbf d_{\rm th}(\theta))^T\,C_{\mathbf d}^{-1}\,(\mathbf d[\alpha_o] - \mathbf d_{\rm th}(\theta)). \label{eq:likelihood}
\end{equation}

In the case of scattering transform, we choose $\mathbf d$ to be the log-reduced first- and second-order coefficients: $\ln s_1^{j_1}$ and $\ln(s_2^{j_1,j_2}/s_1^{j_1})$, where $j_1=1,\cdots,J$, and $j_2=j_1+1,\cdots,J$. Coefficients with $j_2\leq j_1$ are well defined but contain no useful information about the input field, so we do not include them in the analysis. Logarithms of the scattering transform coefficients are chosen because empirically it is found that the distributions of the logarithms are better approximated as multivariate normal distributions~\cite{Cheng_2020}.

To obtain the theoretical summary statistics $\mathbf d_{\rm th}(\theta)$, a large number of realizations of the rotation field $\alpha(\mathbf x)|_\theta$ are generated for a given set of parameters $\theta$ and their summary statistics $\mathbf d[\alpha|_\theta]$ computed. Then,
\begin{equation}
    \mathbf d_{\mathrm{th}}(\theta) = \langle\mathbf d[\alpha|_\theta]\rangle,
\end{equation}
where the average is performed over realizations. In reality, we discretize the parameter space into a finite number of regular grid points $\theta_{\mathbf i}$ and only compute $\mathbf d_{\mathrm{th}}(\theta_{\mathbf i})$ at these grid points. Summary statistics at intermediate points $\theta$ are computed using cubic spline interpolation from the values at the grid points. We check a number of realizations at selected intermediate parameter points $\theta$ to ensure that interpolation errors are well within the standard error due to sample variance.

The set of realizations $\{\alpha(\mathbf x)|_\theta\}$ generated at each point $\theta$ also defines a covariance matrix $C_{\mathbf d}(\theta)$ for the summary statistics $\mathbf d$. For an arbitrary input field, neither the parameters $\theta$ nor the covariance matrix are known \emph{a priori}. Similar to the strategy widely practiced in power spectrum-based analyses in cosmology~\cite{Carron_2013}, we fix a fiducial covariance matrix $C = C(\theta_0)$ at an initial guess $\theta_0$ when evaluating the likelihood function. Once a posterior is obtained, we verify that $\theta_0$ is consistent with the posterior. If not, the likelihood maximization procedure has to be iterated by updating the covariance matrix to $C(\theta_{\mathrm{max}})$, where $\theta_{\mathrm{max}}$ is the parameter set that maximizes the likelihood function in each iteration.

For the $(\mathcal A,\,\xi_0,\,\zeta_0)$ parameter space of the first model described in \refsec{models}, we choose nine grid points for $\xi_0$ evenly spaced between $10^0$ and $10^{1.71}\approx 52$ on the logarithmic scale, and eight grid points of $\zeta_0$ evenly spaced between $10^{-1}$ and $10^0$ again on the logarithmic scale. In the noise-free case, because of the exact scaling behavior $s_1^{j_1}(\mathcal A\,\alpha) = |\mathcal A|\,s_1^{j_1}(\alpha)$ and $s_2^{j_1,j_2}(\mathcal A\,\alpha) = |\mathcal A|\,s_2^{j_1,j_2}(\alpha)$ for any rotation field $\alpha$ (see \refeqs{ST1}{ST2}), we can simply compute the scattering transform coefficients for $\mathcal A=1$ and analytically obtain the likelihood for general values of $\mathcal A$. This eliminates the need to create a grid along the direction of $\mathcal A$ in the parameter space.

For the $(\mathcal A,\,\xi_0,\,f_{\mathrm{sub}})$ parameter space corresponding to string network Model II described in \refsec{models}, we discretize $\xi_0$ as what is done for Model I. We choose eight grid points of $f_{\mathrm{sub}}$ that are evenly spaced between $0.1$ and $1$ on the linear scale.

After $\mathcal A=1$ has been fixed, 2000 realizations of the rotation field in a $128^\circ \times 128^\circ$ patch of sky are generated at a $1024\times 1024$ pixel resolution, at each grid point in either the $(\xi_0,\,\zeta_0)$-space or the $(\xi_0,\,f_{\mathrm{sub}})$-space, and according to the procedure outlined in \refsec{realization}. This is a computationally expensive step, which involves painting $10^3$ to $10^6$ ellipses per realization. Applying scattering transform to each of the realizations, we obtain the statistics of the scattering transform coefficients of the noise-free rotation field at each grid point in the parameter space.

For the case in which instrumental and reconstruction noise is taken into account, it is necessary to introduce grid points to sample a range of signal strengths ($\mathcal A^2\,\xi_0$) in comparison to the noise. By a simple rescaling, each noise-free realization of the rotation signal generated with $\mathcal A=1$ now becomes sixteen noise-free realizations with $\mathcal A^2\,\xi_0$ evenly spaced between $10^{-3}$ and $10^{2.5}$ on the logarithmic scale. This is done twice for each noise-free realization and each signal strength $\mathcal A^2\,\xi_0$. Unique realizations of the reconstruction noise (corresponding to CMB-HD noise level) are then generated and added to each of the noise-free realizations of the signal. The scattering transform calculation is then performed for each noisy rotation field realization.

Scattering transform for large two-dimensional pixelated images is in general computationally expensive. Therefore, we down-sample each $1024\times 1024$-pixel map to a $128\times 128$-pixel map by filtering out Fourier modes of high spatial frequencies. This removes information that is potentially dependent on the aliasing properties of the pixelated ellipses, while ensuring that the rotation power spectrum of each map is preserved at low spatial frequencies.

We use the implementation of scattering transform in the \textsc{Python} package \textsc{Kymatio}~\cite{JMLR:v21:19-047} and compute coefficients for $L=8$ azimuthal orientations and $J=6$ logarithmically spaced dilation scales. This results in 6 first-order and 15 second-order reduced scattering transform coefficients.

To demonstrate parameter inference, we generate a mock signal with known model parameters $\theta_{\mathrm{mock}}$, and then use the Markov Chain Monte Carlo (MCMC) method with the log-likelihood function \refeq{likelihood} to estimate the posterior distribution of $\theta$. The fiducial covariance matrix $C_{\mathbf d}$ is chosen to be $C(\theta_{\mathrm{mock}})$. In each MCMC run, 32 parallel walkers are used, taking $10^4$ steps each. The first $100$ samples in each chain are discarded.

Flat priors are chosen for the logarithmic quantities $\log_{10}\mathcal A$, $\log_{10}\xi_0$, and $\log_{10}\zeta_0$, and for the linear quantity $f_{\mathrm{sub}}$. The priors are defined within ranges that are constrained by the ranges of parameters used to generate realizations for the theoretical scattering transform coefficients. These are summarized in \reftab{priors}.

\begin{table}[h!]
    \centering
    \begin{tabular}{c|c|c|c}
        \hline
        \hline
         & $\log_{10}(\mathcal A^2\,\xi_0)$ & $\log_{10}\xi_0$ & $\log_{10}\zeta_0$ or $f_{\mathrm{sub}}$ \\
        \hline
        Uniform radius, noise-free & $-10\sim 10$ & $0\sim 1.71$ & $-1\sim 0$ \\
        Uniform radius, noisy & $-3\sim 2.5$ & $0\sim 1.71$ & $-1\sim 0$ \\
        Log-distributed radii, noise-free & $-10\sim 10$ & $0\sim 1.71$ & $0.1\sim 1$ \\
        Log-distributed radii, noisy & $-3\sim 2.5$ & $0\sim 1.71$ & $0.1\sim 1$ \\
        \hline
        \hline
    \end{tabular}
    \caption{Ranges for (log-)flat priors used in posterior estimation.}
    \label{tab:priors}
\end{table}

To compare between the rotation power spectrum and scattering transform coefficients as summary statistics, the aforementioned likelihood maximization procedure is also repeated with $\mathbf d$ chosen to be the logarithm of 20 binned power spectrum points between $L=1$ and $L=240$. Finally, joint inference is performed by combining the binned power spectrum points and the scattering transform coefficients as a single set of summary statistics. Taking into account the information from both types of summary statistics is expected to improve the parameter inference results.

\section{Results}
\label{sec:results}

\subsection{Noise-free case}
\label{sec:results_noise_free}
We first study the constraining power of scattering transform on the value of $\mathcal A$ for the loop-crossing model in the noise-free case. This corresponds to an ideal ``best-case'' scenario in which the the impact of polarization measurement errors, foreground contamination, and the statistical variance in the reconstruction of the rotation angle is negligibly small.

If a single cosmic string in the sky could be resolved angularly, then $\mathcal A$ could be obtained directly by measuring the difference in CMB polarization angle on both sides of the string. The finite angular resolution achievable for the rotation field is the primary limitation in the measurement of $\mathcal A$ from the polarization rotation field, even in the noise-free case. The angular resolution of rotation field used here is $\sfrac{1}{8}$ degrees, corresponding to multipoles up to $L=1440$.

\begin{figure}
    \centering
    \begin{subfigure}{0.48\textwidth}
        \centering
        \includegraphics[width=\textwidth]{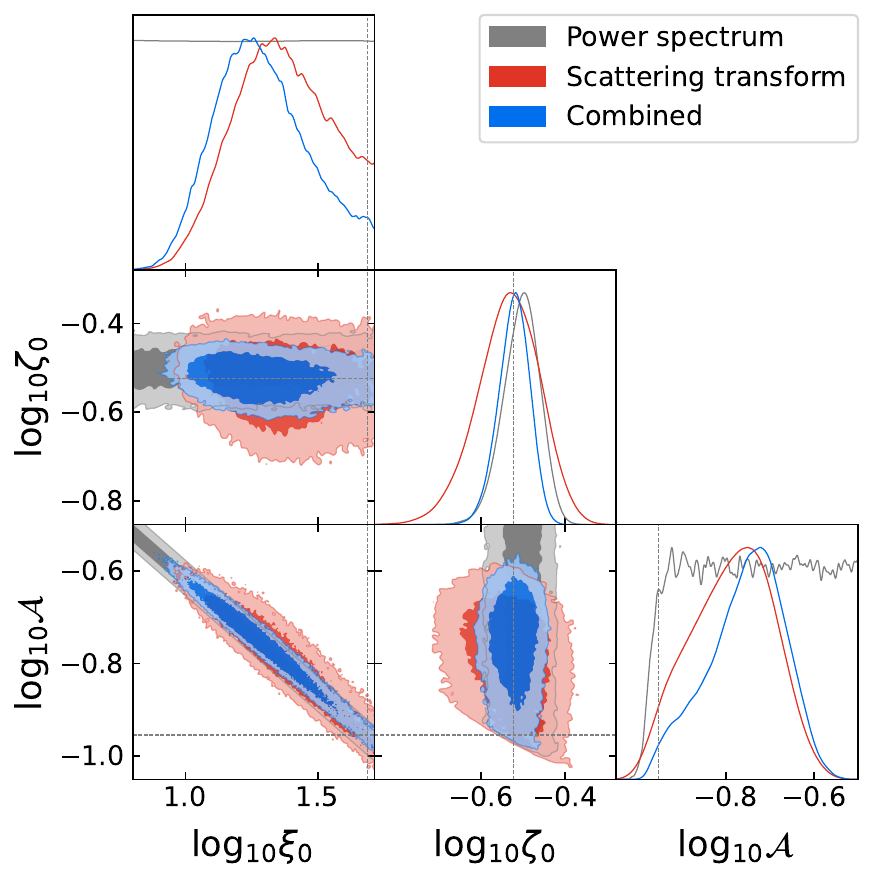}
        \caption{$\mathcal A^2\xi_0 = 0.6,\ \mathcal A = \sfrac{1}{9},\ \zeta_0 = 0.3$}
    \end{subfigure}
    \begin{subfigure}{0.48\textwidth}
        \centering
        \includegraphics[width=\textwidth]{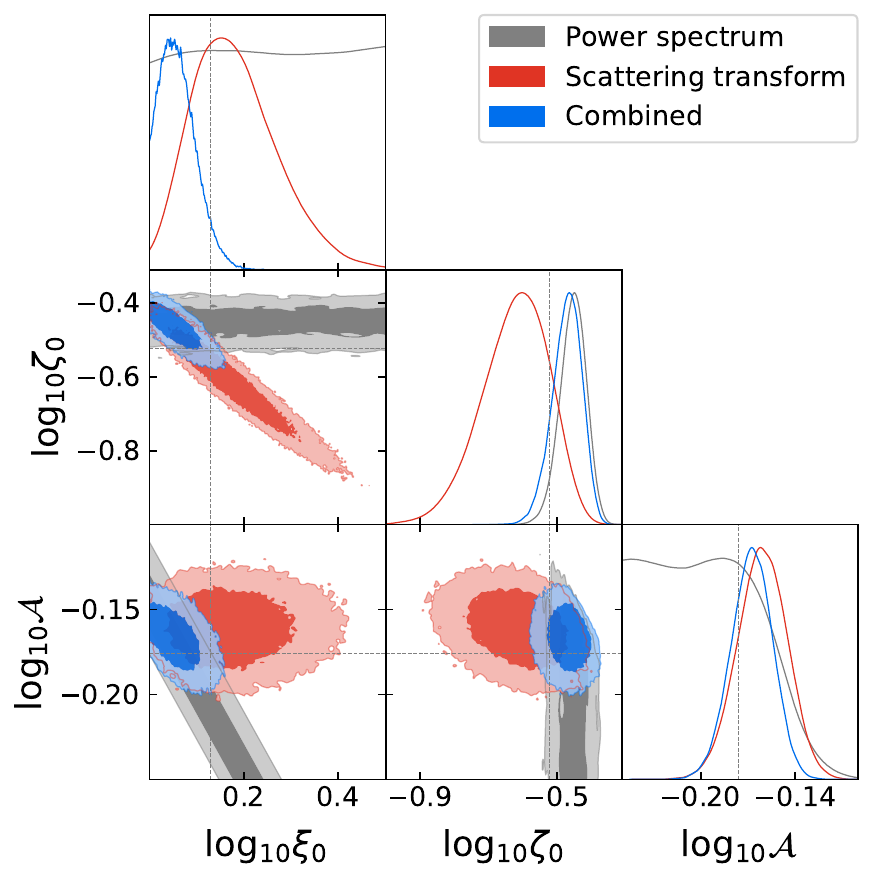}
        \caption{$\mathcal A^2\xi_0 = 0.6,\ \mathcal A = \sfrac{2}{3},\ \zeta_0 = 0.3$}
    \end{subfigure}
    \par\bigskip
    \begin{subfigure}{0.48\textwidth}
        \centering
        \includegraphics[width=\textwidth]{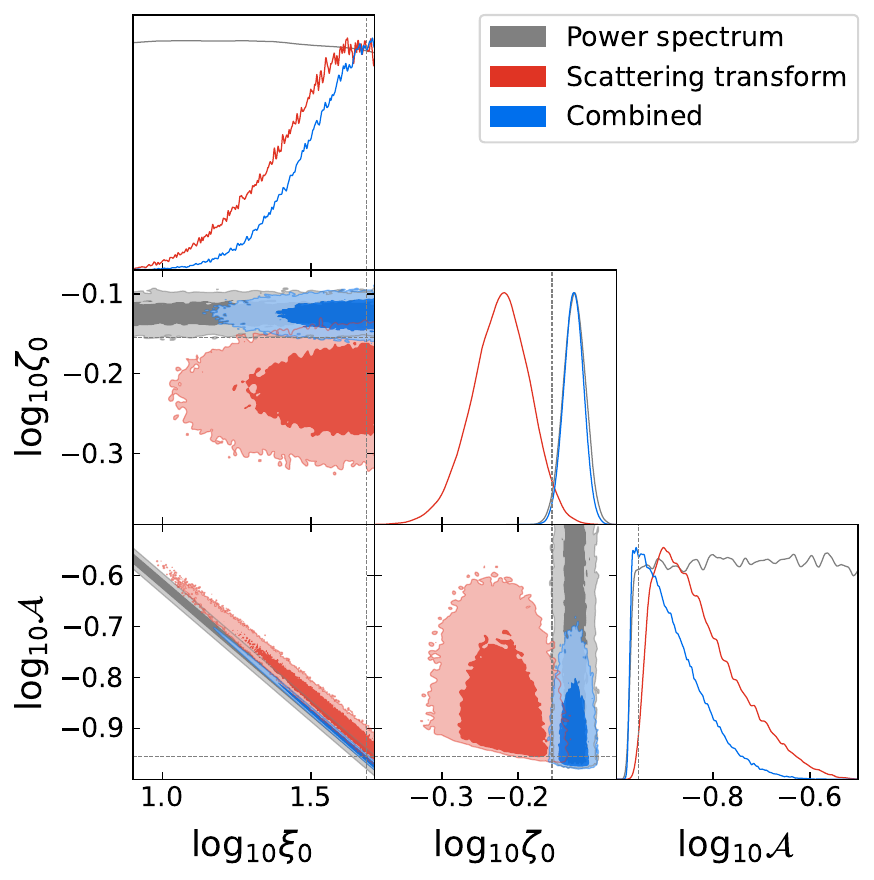}
        \caption{$\mathcal A^2\xi_0 = 0.6,\ \mathcal A = \sfrac{1}{9},\ \zeta_0 = 0.7$}
    \end{subfigure}
    \begin{subfigure}{0.48\textwidth}
        \centering
        \includegraphics[width=\textwidth]{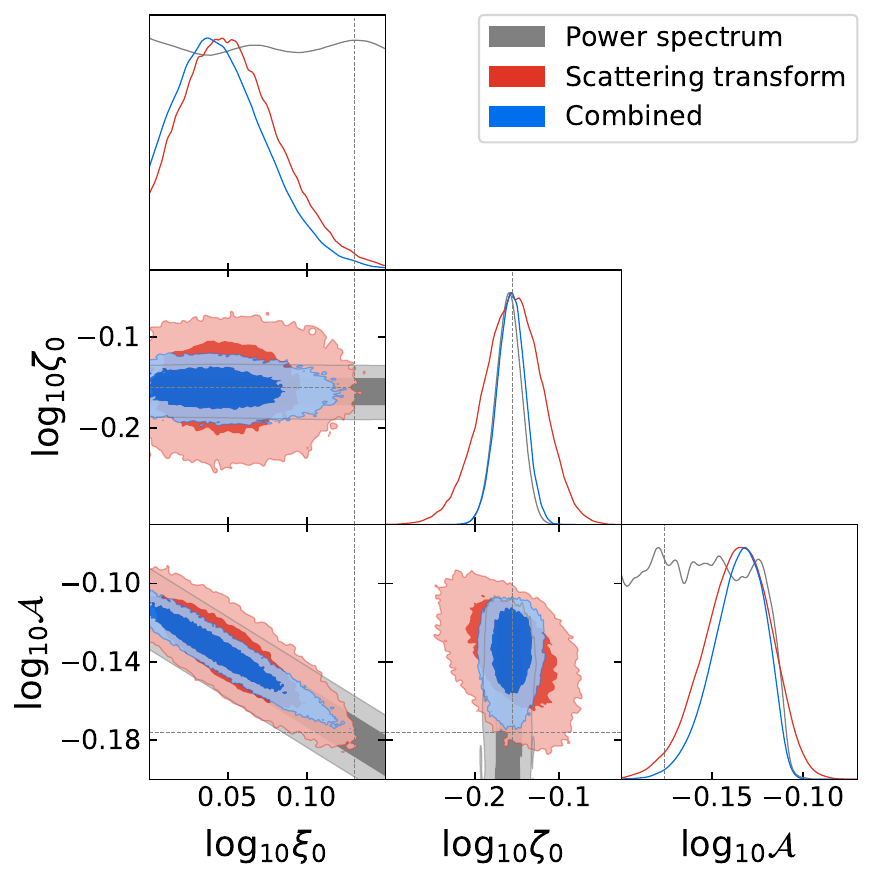}
        \caption{$\mathcal A^2\xi_0 = 0.6,\ \mathcal A = \sfrac{2}{3},\ \zeta_0 = 0.7$}
    \end{subfigure}
    \caption{Posterior distributions for the parameters of the loop-crossing string network model with a unique string loop radius (Model I). Inference is performed in the limit of negligible reconstruction noise for the polarization rotation. Three sets of summary statistics are exploited: rotation power spectrum (grey), scattering transform coefficients (red), and the combination of the two (blue). True parameter values used for the input maps are marked with grey lines and given in the captions.}
    \label{fig:triangles_model1_noiseless}
\end{figure}

\begin{figure}
    \centering
    \begin{subfigure}{0.48\textwidth}
        \centering
        \includegraphics[width=\textwidth]{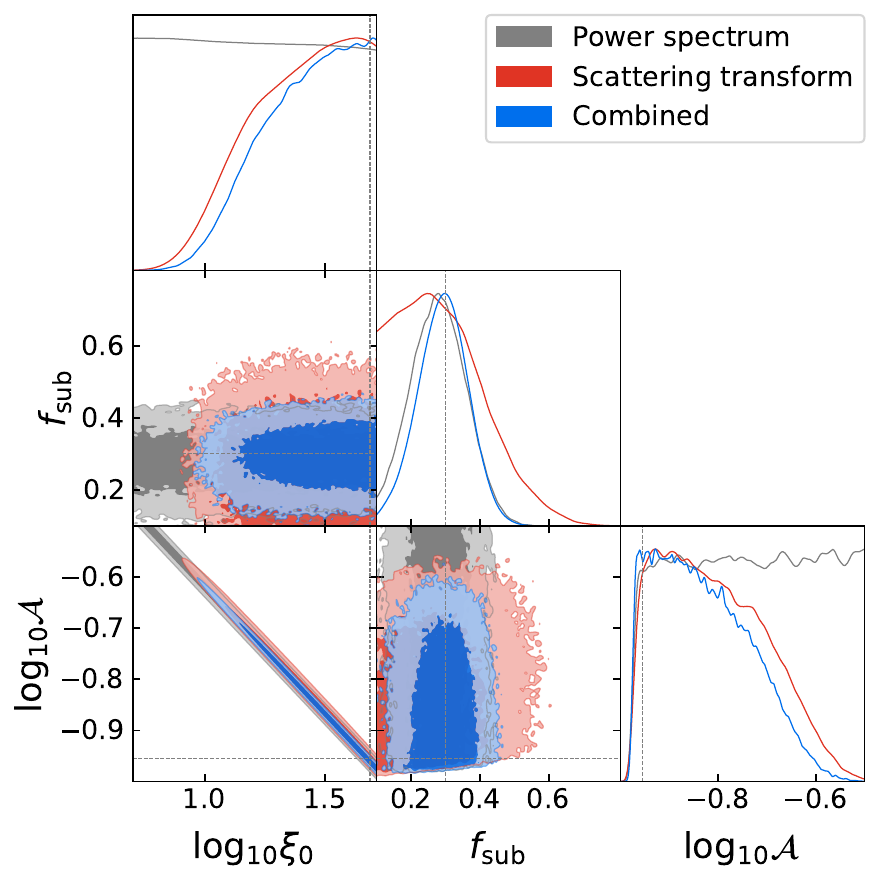}
        \caption{$\mathcal A^2\xi_0 = 0.6,\ \mathcal A = \sfrac{1}{9},\ f_{\mathrm{sub}} = 0.3$}
    \end{subfigure}
    \begin{subfigure}{0.48\textwidth}
        \centering
        \includegraphics[width=\textwidth]{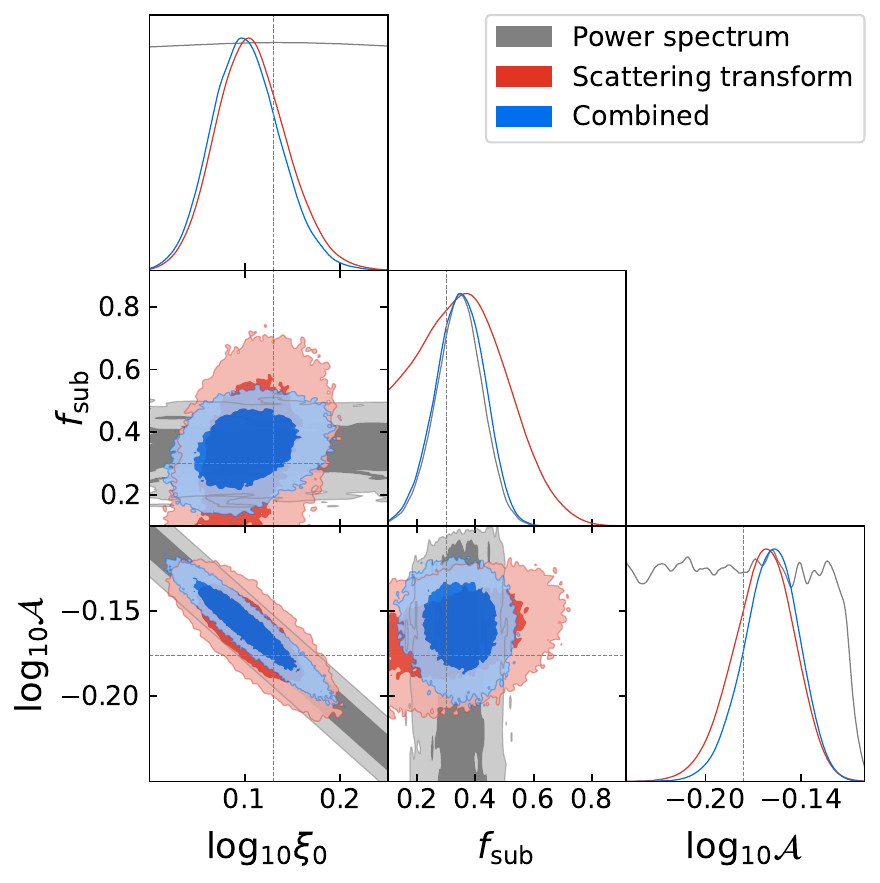}
        \caption{$\mathcal A^2\xi_0 = 0.6,\ \mathcal A = \sfrac{2}{3},\ f_{\mathrm{sub}} = 0.3$}
    \end{subfigure}
    \par\bigskip
    \begin{subfigure}{0.48\textwidth}
        \centering
        \includegraphics[width=\textwidth]{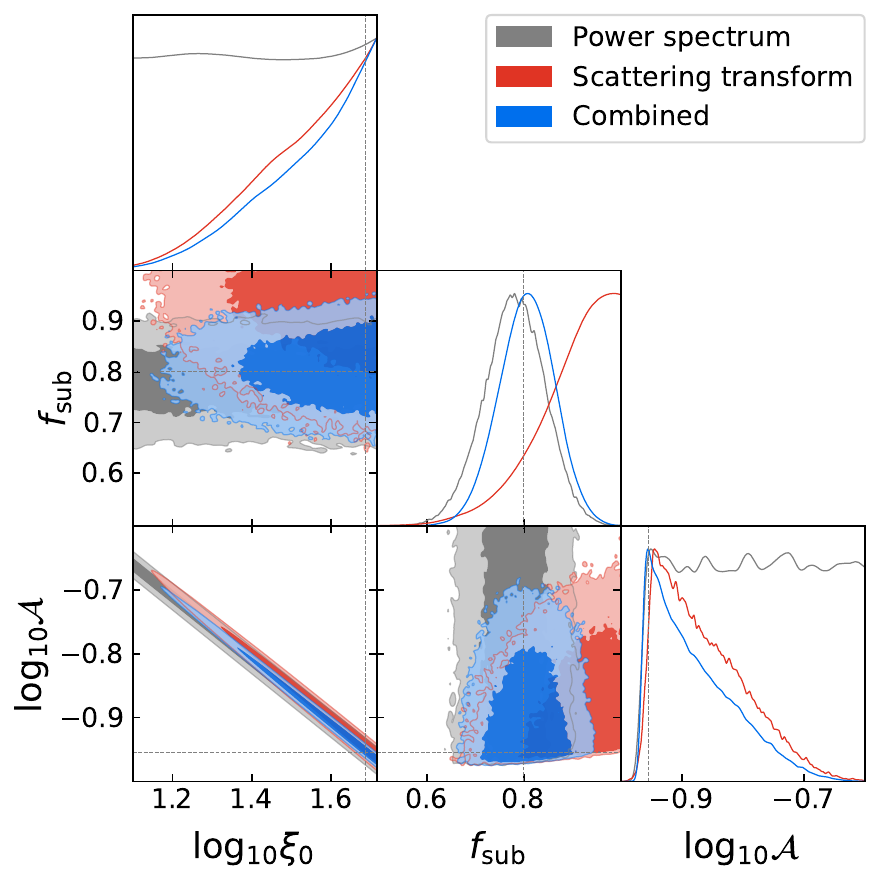}
        \caption{$\mathcal A^2\xi_0 = 0.6,\ \mathcal A = \sfrac{1}{9},\ f_{\mathrm{sub}} = 0.8$}
    \end{subfigure}
    \begin{subfigure}{0.48\textwidth}
        \centering
        \includegraphics[width=\textwidth]{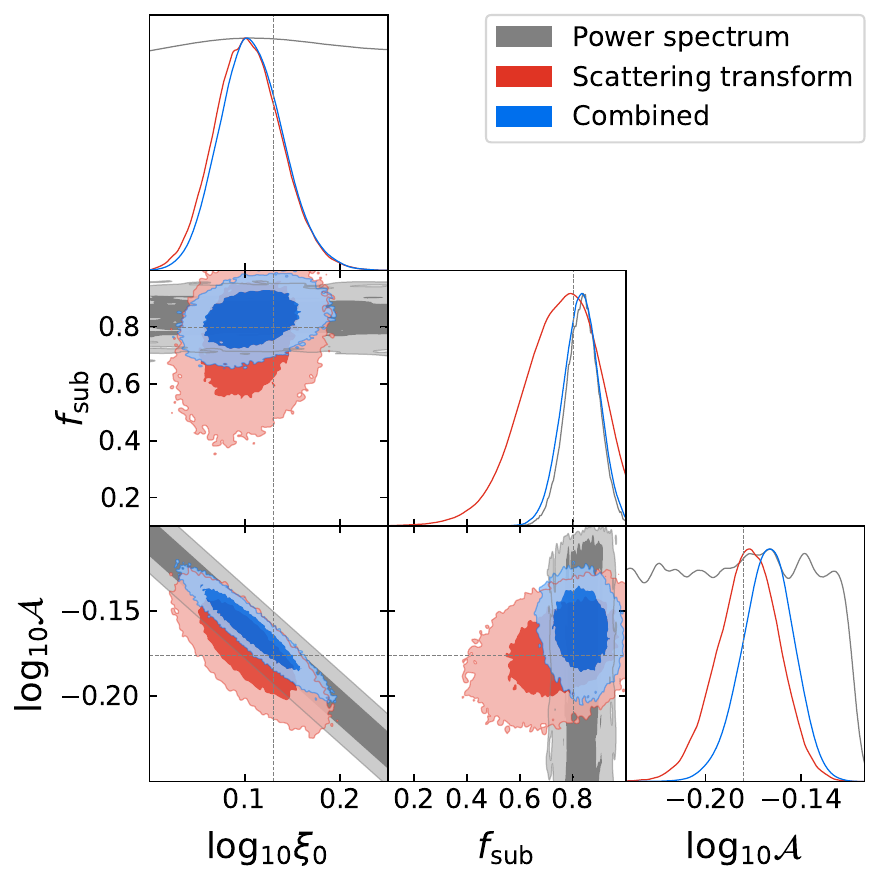}
        \caption{$\mathcal A^2\xi_0 = 0.6,\ \mathcal A = \sfrac{2}{3},\ f_{\mathrm{sub}} = 0.8$}
    \end{subfigure}
    \caption{Posterior distributions for the parameters of the loop-crossing string network model with log-flat distributed string loop radii (Model II). Inference is performed in the limit of negligible reconstruction noise for the polarization rotation.}
    \label{fig:triangles_model2_noiseless}
\end{figure}

Noise-free rotation realizations are independently generated according to the method explained in \refsec{realization} and are treated as mock input fields for the inference procedure of \refsec{method}. The posteriors of $(\mathcal A,\,\xi_0,\,\zeta_0)$ estimated from the mock input fields using MCMC on the likelihood in \refeq{likelihood} are presented in \reffig{triangles_model1_noiseless} for Model I and \reffig{triangles_model2_noiseless} for Model II for select parameters. Posteriors obtained from the same realization by power spectrum and scattering transform are consistent with each other as well as the true parameter values. The $\mathcal A^2\,\xi_0$ parameter degeneracy in the power spectrum inference (posteriors in grey) is apparent, and the 1D posteriors for $\mathcal A$ and $\xi_0$ individually are essentially flat within the bounds set by our adopted priors (\reftab{priors}). In all cases, scattering transform successfully breaks this degeneracy by reducing the width of the posterior along the degenerate direction to a finite value. In some cases, sharp straight edges in the contours of the 2D joint distributions artificially arise as a result of the boundaries of the adopted priors. \reffig{triangles_model1_noiseless} shows that parameter degeneracy is most significantly diminished for lower $\xi_0$ (and hence higher $\mathcal A$ at fixed $\mathcal{A}^2\,\xi_0$) and lower $\zeta_0$, corresponding to rotation fields with a higher degree of non-Gaussianity.

We notice that the scattering transform method results in less tightly constrained posteriors in the direction perpendicular to the $\mathcal{A}^2\,\xi_0$ degeneracy. We attribute this to the fact that the first-order scattering transform coefficients, while conceptually analogous to binned power spectrum, have a poorer resolution in Fourier space (6 vs.~20 points). Significantly improved inference results can be achieved by trivially combining binned power spectrum and scattering transform coefficients into an overall set of summary statistics, whose resultant posteriors are shown in blue in the figures.

\begin{figure}
    \begin{center}
        \scalebox{0.7}{\input{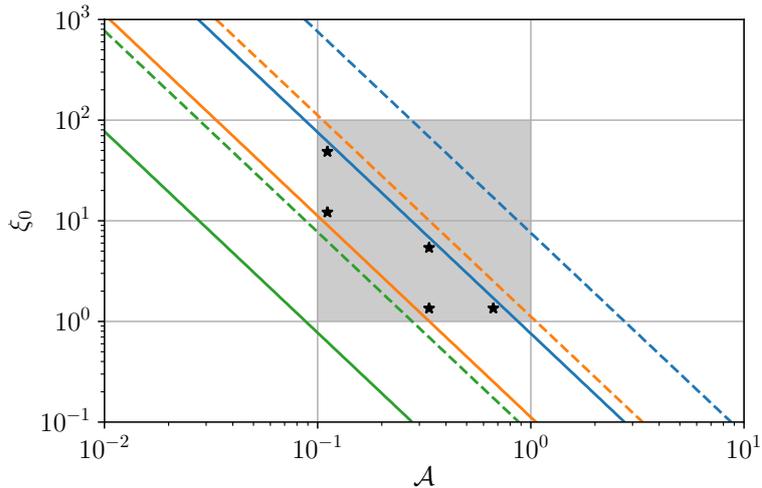}}
    \end{center}
    \caption{The $(\mathcal A,\,\xi_0)$ parameter space. The power spectrum of the rotation field is unchanged along the degenerate direction---from upper-left to lower-right. Stars mark the five points---three along $\mathcal A^2\,\xi_0 = 0.6$ and two along $\mathcal A^2\,\xi_0 = 0.15$---chosen for the mock input rotation fields from which $\mathcal A$ will be estimated. Colored lines mark the $3\sigma$ (solid) and $30\sigma$ (dashed) detectability of a string network described by Model II ($f_{\mathrm{sub}} = 0.6$), for three CMB experiments: Planck SMICA (blue), Simons Observatory (orange), and CMB-S4 (green). The region of parameter space of the most theoretical interest is shaded in grey~\cite{Agrawal_2020}.}
    \label{fig:param_space_points}
\end{figure}

Our goal is to distinguish realizations with different values of $\mathcal A$ while keeping the combination $\mathcal A^2\,\xi_0$ constant. For $\mathcal A^2\,\xi_0 = 0.6$, which is just below the current Planck constraint, we consider mock rotation fields with $\mathcal A = \sfrac{1}{9},\,\sfrac{1}{3},\,\sfrac{2}{3}$, [comments on which UV models give these values]. For $\mathcal A^2\xi_0 = 0.15$, which Simons Observatory would be able to detect at a $3\sigma$ level, we consider $\mathcal A=\sfrac{1}{9},\,\sfrac{1}{3}$. For each of these five choices of parameters, we also consider $\zeta_0 = 0.3,\,0.7$ and $f_{\mathrm{sub}} = 0.3,\,0.8$ for the two loop-crossing models, respectively. \reffig{param_space_points} marks these points in the parameter space, which can be compared to the theoretically compelling region of the parameter space in grey, as well as to the regions of detectability by several representative CMB experiments.

\begin{figure}
    \centering
    \begin{subfigure}{0.48\textwidth}
        \centering
        \includegraphics[width=\textwidth]{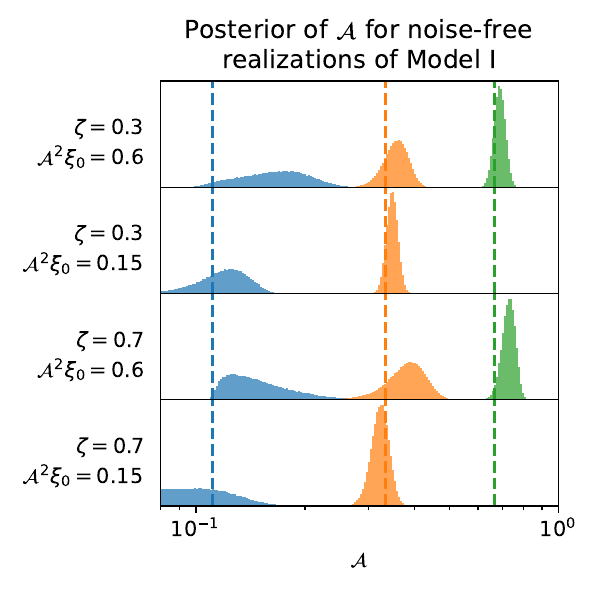}
    \end{subfigure}
    \begin{subfigure}{0.48\textwidth}
        \centering
        \includegraphics[width=\textwidth]{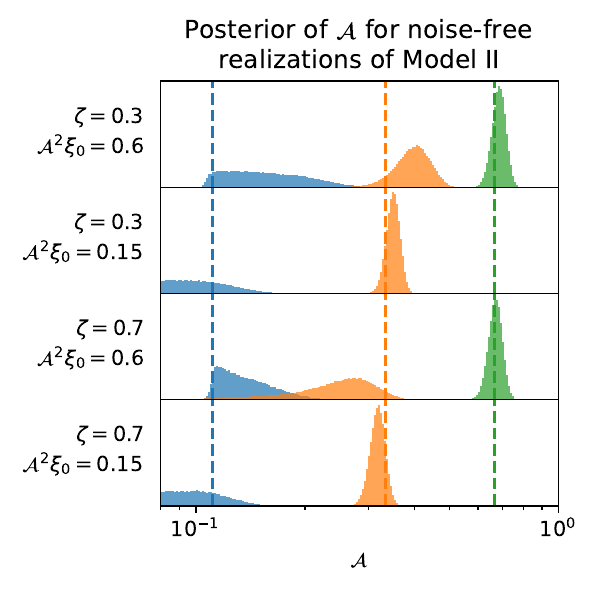}
    \end{subfigure}
    \caption{1D posteriors of $\mathcal A$ obtained from noise-free realizations of the loop-crossing model using scattering transform as the summary statistics. This demonstrates the best-case scenario of the estimation of $\mathcal A$ for the chosen parameters in \reffig{param_space_points}. Vertical lines mark the true values $\mathcal A=\sfrac{1}{9},\,\sfrac{1}{3},\,\sfrac{2}{3}$ of the corresponding input fields.}
    \label{fig:posteriors_noiseless}
\end{figure}

The 1D posteriors of $\mathcal A$ obtained from realizations of the chosen sets of parameters using scattering transform as the summary statistics are summarized in \reffig{posteriors_noiseless}. In this demonstration of the best-case scenario, realizations with different values $\mathcal A=\sfrac{1}{9},\,\sfrac{1}{3},\,\sfrac{2}{3}$ can all be clearly distinguished from each other. Therefore, scattering transform shows dramatic improvement over the power spectrum method, which produces completely flat posteriors of $\mathcal A$ (up to the boundaries set by the prior) due to the $\mathcal A^2\,\xi_0$ degeneracy. We also observe that $\mathcal A=\sfrac{1}{9},\,\sfrac{1}{3}$ can be better distinguished for a lower value of $\mathcal A^2\,\xi_0$, as a result of a lower $\xi_0$ and hence stronger non-Gaussian features in the rotation field, which scattering transform is able to exploit.

\subsection{Noisy case}
We now examine the parameter inference power of the scattering transform method for the loop-crossing model, taking into account realistic reconstruction noise for the rotation field.

The forecast reconstruction noise of CMB-S4 is low enough that the rotation signal of an axion string network described by the loop-crossing model can be either confirmed or falsified within the theoretically favored region of the parameter space~\cite{Yin_2022}. Should such a signal be discovered in experiments, it will be imperative to quantify it in more detail to distinguish between patterns that resemble Gaussian random fields and patterns that arise from an axion string network. It will also be important to measure the value of the electromagnetic anomaly coefficient $\mathcal A$ to gain insight into the charge quantization in beyond-SM theories. Although CMB-S4 will be capable of statistically detecting the rotation signal, an even more sensitive experiment will be needed to adequately break the $\mathcal A^2\,\xi_0$ parameter degeneracy at the power spectrum level and reveal the non-Gaussian nature of the rotation pattern in the sky. For demonstration, we set experimental parameters to be those of the proposed concept CMB-HD: $\Delta_T = 0.5\,\mathrm{\mu K\,arcmin}$, $\Delta_P = \sqrt{2}\,\Delta_T$, and $\Theta_{FWHM} = 0.25\,\mathrm{arcmin}$ \cite{Ferraro_2022}.

\begin{figure}
    \centering
    \begin{subfigure}{0.48\textwidth}
        \centering
        \includegraphics[width=\textwidth]{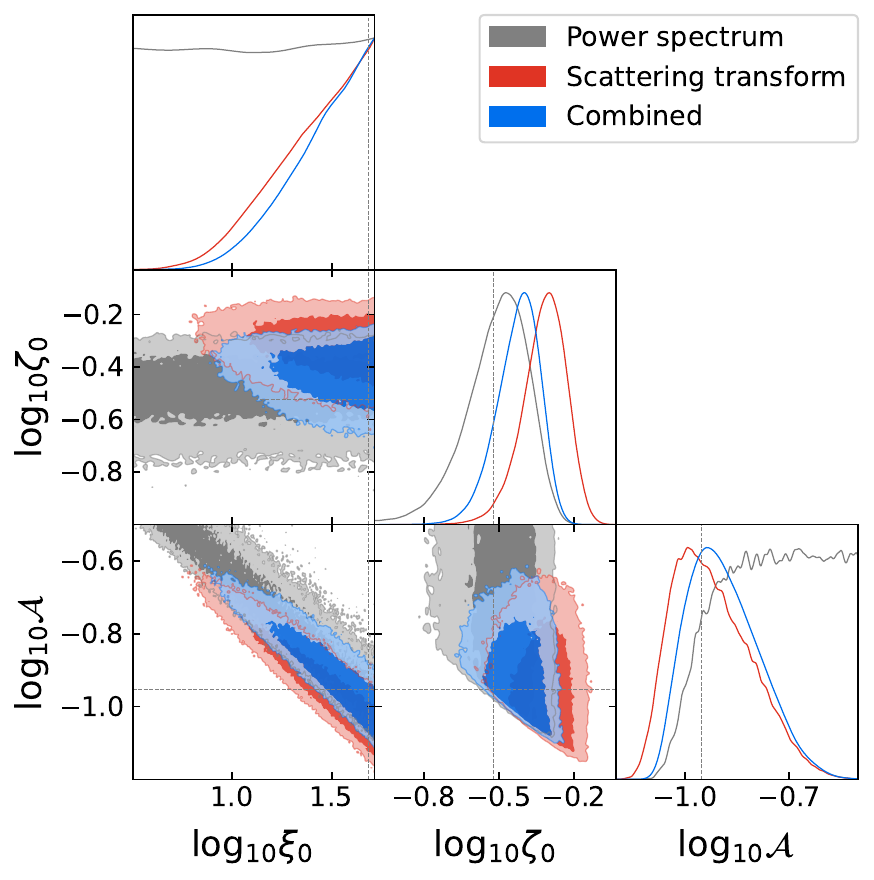}
        \caption{$\mathcal A^2\xi_0 = 0.6,\ \mathcal A = \sfrac{1}{9},\ \zeta_0 = 0.3$}
    \end{subfigure}
    \begin{subfigure}{0.48\textwidth}
        \centering
        \includegraphics[width=\textwidth]{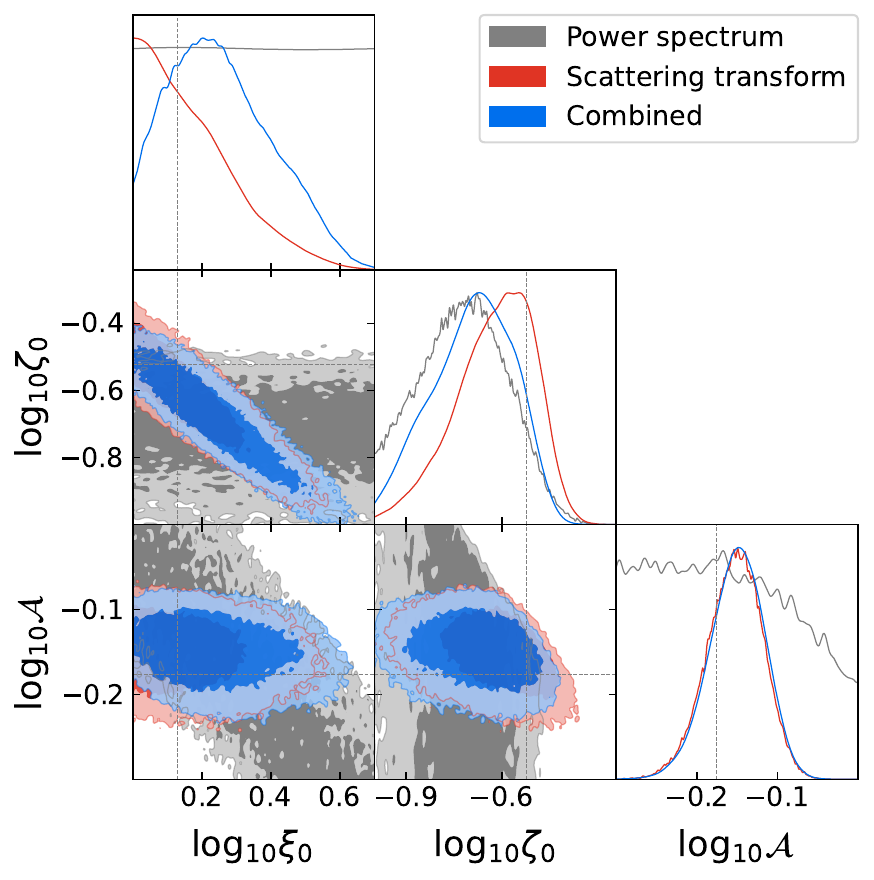}
        \caption{$\mathcal A^2\xi_0 = 0.6,\ \mathcal A = \sfrac{2}{3},\ \zeta_0 = 0.3$}
    \end{subfigure}
    \par\bigskip
    \begin{subfigure}{0.48\textwidth}
        \centering
        \includegraphics[width=\textwidth]{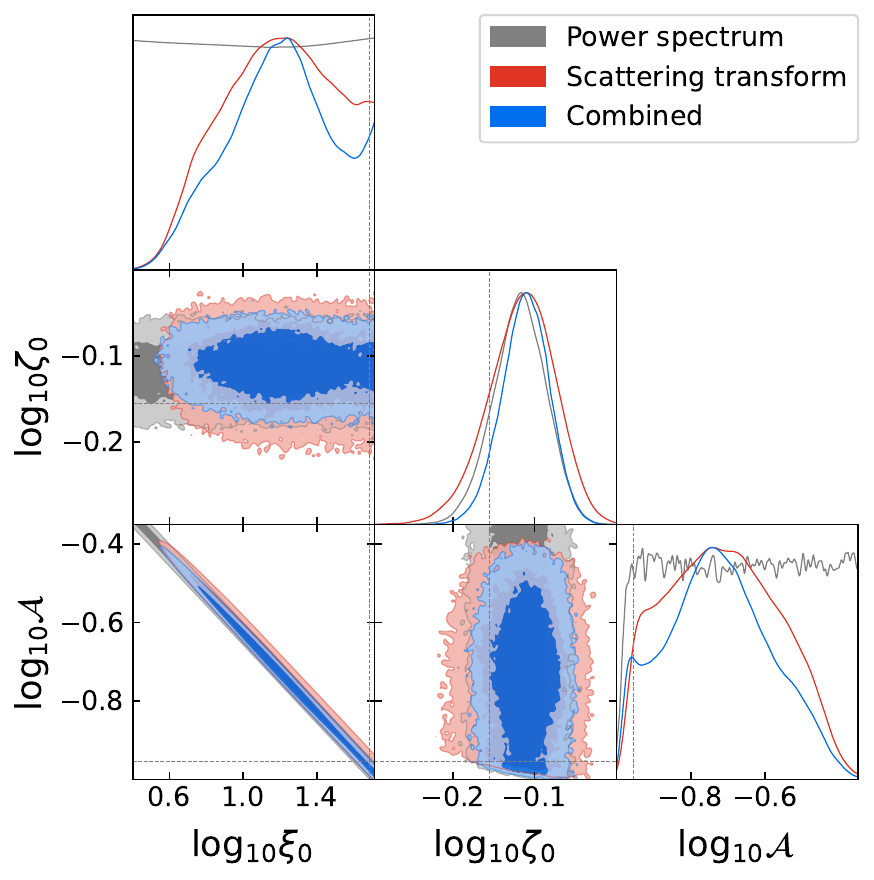}
        \caption{$\mathcal A^2\xi_0 = 0.6,\ \mathcal A = \sfrac{1}{9},\ \zeta_0 = 0.7$}
    \end{subfigure}
    \begin{subfigure}{0.48\textwidth}
        \centering
        \includegraphics[width=\textwidth]{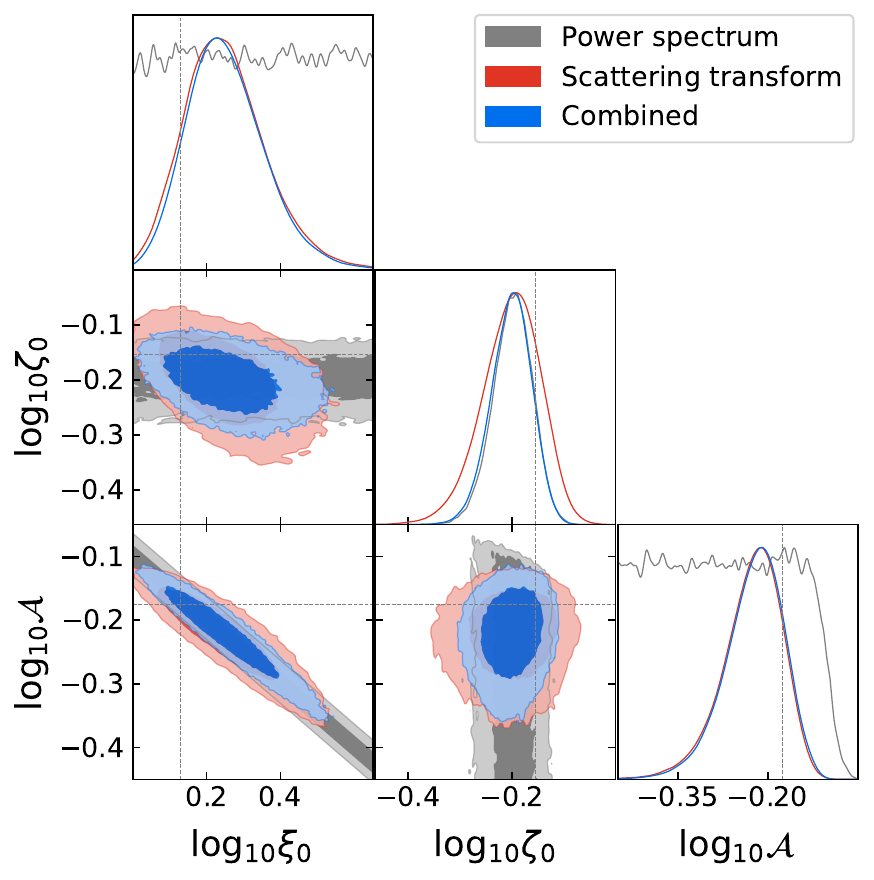}
        \caption{$\mathcal A^2\xi_0 = 0.6,\ \mathcal A = \sfrac{2}{3},\ \zeta_0 = 0.7$}
    \end{subfigure}
    \caption{Posterior distributions for the parameters of the loop-crossing string network model with a unique string loop radius (Model I). Inference is performed at the CMB-HD reconstruction noise level.}
    \label{fig:triangles_model1_cmbhd}
\end{figure}

\begin{figure}
    \centering
    \begin{subfigure}{0.48\textwidth}
        \centering
        \includegraphics[width=\textwidth]{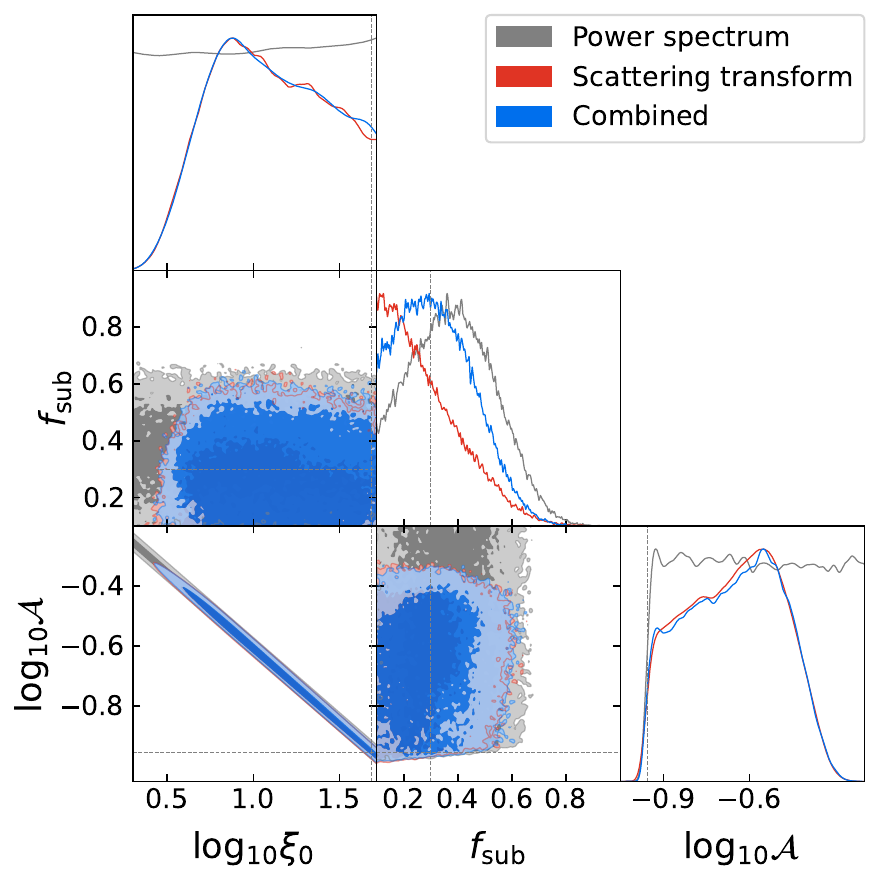}
        \caption{$\mathcal A^2\xi_0 = 0.6,\ \mathcal A = \sfrac{1}{9},\ f_{\mathrm{sub}} = 0.3$}
    \end{subfigure}
    \begin{subfigure}{0.48\textwidth}
        \centering
        \includegraphics[width=\textwidth]{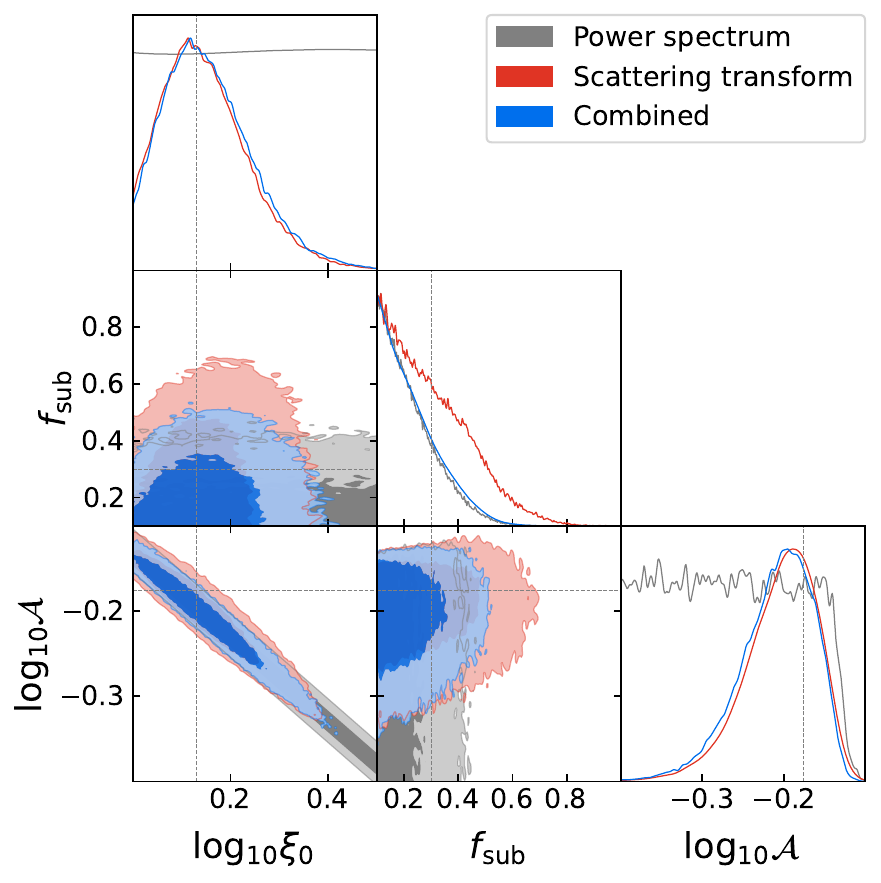}
        \caption{$\mathcal A^2\xi_0 = 0.6,\ \mathcal A = \sfrac{2}{3},\ f_{\mathrm{sub}} = 0.3$}
    \end{subfigure}
    \par\bigskip
    \begin{subfigure}{0.48\textwidth}
        \centering
        \includegraphics[width=\textwidth]{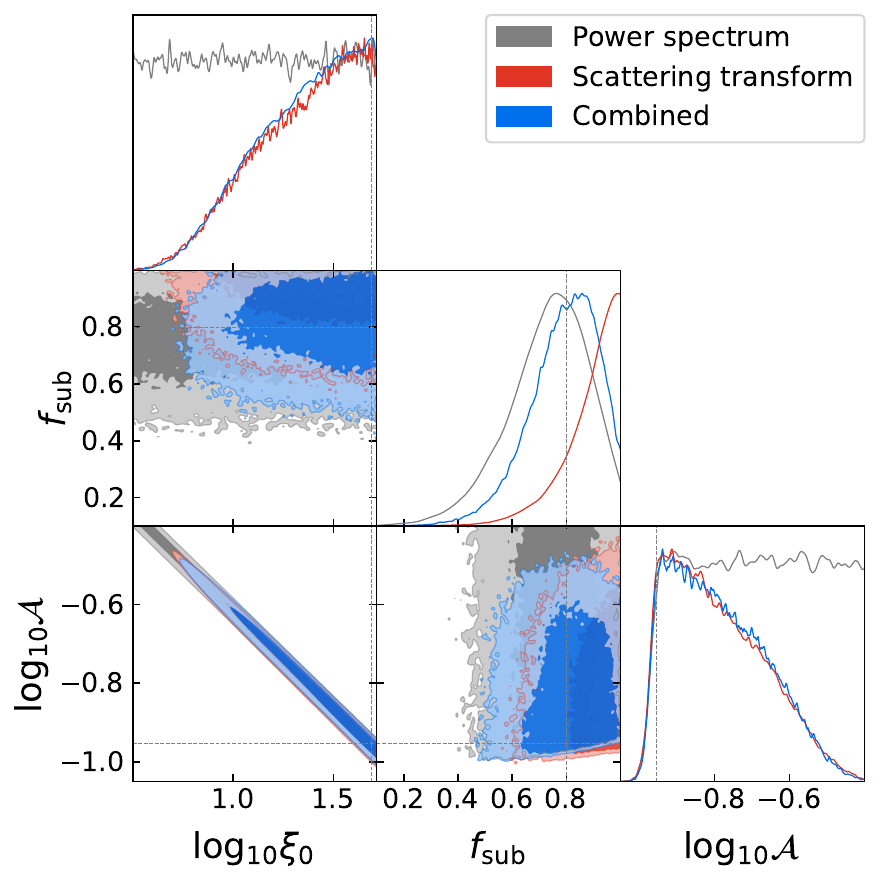}
        \caption{$\mathcal A^2\xi_0 = 0.6,\ \mathcal A = \sfrac{1}{9},\ f_{\mathrm{sub}} = 0.8$}
    \end{subfigure}
    \begin{subfigure}{0.48\textwidth}
        \centering
        \includegraphics[width=\textwidth]{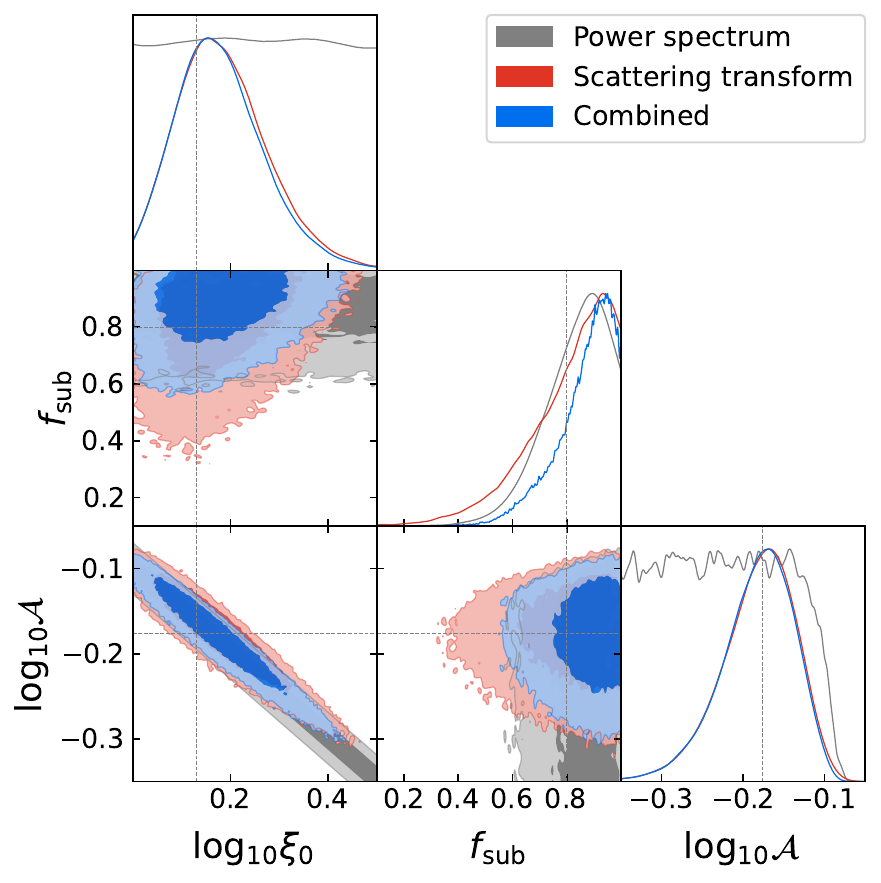}
        \caption{$\mathcal A^2\xi_0 = 0.6,\ \mathcal A = \sfrac{2}{3},\ f_{\mathrm{sub}} = 0.8$}
    \end{subfigure}
    \caption{Posterior distributions for the parameters of the loop-crossing string network model with log-flat distributed string loop radii (Model II). Inference is performed at the CMB-HD reconstruction noise level.}
    \label{fig:triangles_model2_cmbhd}
\end{figure}

Realizations of the rotation field with reconstruction noise injected are independently generated according to the procedures explained in \refsec{realization} and \refsec{noise}. They are treated as mock input to the parameter inference procedure of \refsec{method}. The posteriors of $(\mathcal A,\,\xi_0,\,\zeta_0)$ estimated from the mock input fields using MCMC on the likelihood in \refeq{likelihood} are presented in \reffig{triangles_model1_cmbhd} for Model I and \reffig{triangles_model2_cmbhd} for Model II for select parameters. The joint posteriors show the same qualitative features as the ones in the noise-free case (\refsec{results_noise_free}), albeit less tightly constrained.

\begin{figure}
    \centering
    \begin{subfigure}{0.48\textwidth}
        \centering
        \includegraphics[width=\textwidth]{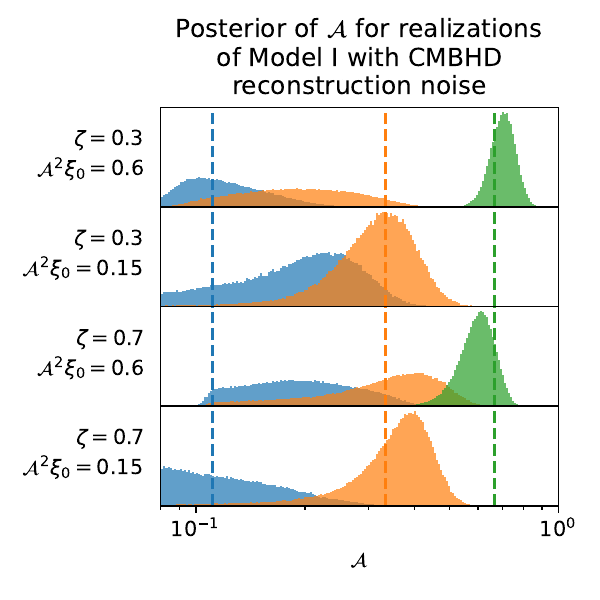}
    \end{subfigure}
    \begin{subfigure}{0.48\textwidth}
        \centering
        \includegraphics[width=\textwidth]{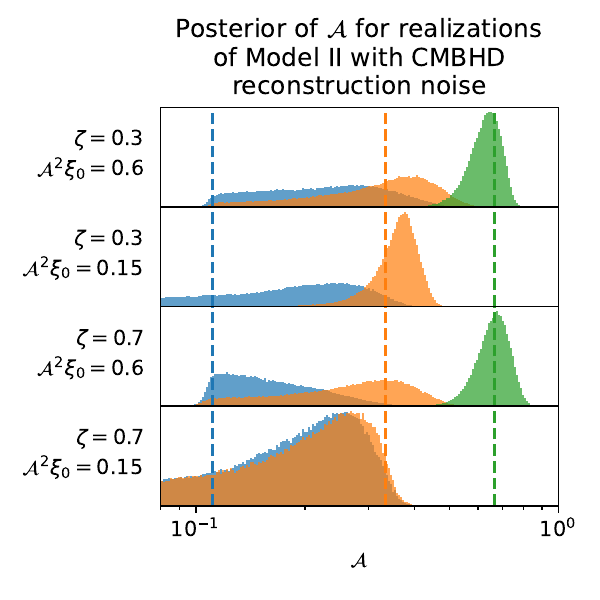}
    \end{subfigure}
    \caption{1D posteriors of $\mathcal A$ obtained from realizations of the loop-crossing model with CMB-HD reconstruction noise level using scattering transform as the summary statistics. This demonstrates that CMB-HD will be able to constrain beyond-SM theories by constraining $\mathcal A$ from a string network detectable at the Simons Observatory. Vertical lines mark the true values $\mathcal A=\sfrac{1}{9},\,\sfrac{1}{3},\,\sfrac{2}{3}$ of the corresponding input fields.}
    \label{fig:posteriors_cmbhd}
\end{figure}

For the same choices of 10 sets of parameters each for Model I and II, we present in \reffig{posteriors_cmbhd} the 1D posteriors of $\mathcal A$ obtained from realizations with CMB-HD reconstruction noise level using scattering transform as the summary statistics. Although the posteriors are wider than in \reffig{posteriors_noiseless}, they nevertheless demonstrate that CMB-HD will be able to clearly distinguish $\mathcal A=\sfrac{2}{3}$ from $\sfrac{1}{9}$ and $\sfrac{1}{3}$ and marginally distinguish $\mathcal A=\sfrac{1}{9}$ from $\sfrac{1}{3}$, by exploiting the non-Gaussian information in the rotation field of a string network with $\mathcal A^2\,\xi_0 = 0.6$, just below the sensitivity of Planck SMICA. We conclude that the scattering transform of string-induced cosmic birefringence can provide a strong test of charge quantization in beyond-SM physics.

\section{Discussion}\label{sec:discussion}

The contrast in the first rows of \reffig{triangles_model1_cmbhd} and \reffig{triangles_model2_cmbhd} shows that the inference power of various summary statistics also depends on the choice of phenomenological model of cosmic string networks. Since scattering transform excels by extracting non-Gaussian information from the input field, its improvement over power spectrum is greatest when the chosen phenomenological model produces strongly non-Gaussian features, such as large patches with coherent polarization rotation.

The loop-crossing model has the advantage of being easy to compute, but in any realization of it (for any distribution of radii), it produces a large number of angularly small string loops at high redshift. The aggregate of these small string loops is similar to Poisson shot noise, which for high $\xi_0$ contributes a Gaussian random field. The inference power of scattering transform may be significantly hampered by this specific feature of the loop-crossing model.

In contrast, one can imagine an alternative phenomenological model of cosmic string networks in which circular arc segments are rearranged into irregular non-circular loops while preserving the overall distribution of string length as a function of radius of curvature. This mimics the more realistic situation in which strings loops are not perfect circles, but have ``wiggles'' that result from string network dynamics. By keeping the same $\xi_0$ and therefore the same mean energy density, such a phenomenological model would produce fewer, hence larger, patches of coherent rotation. This corresponds to higher non-Gaussian information in the rotation field, so the best-case estimates of $\mathcal A$ using scattering transform are expected to be even more stringent than those shown in \reffig{posteriors_noiseless}. Compared to string networks realized in physical simulations, the loop-crossing model perhaps represents a pessimistic scenario regarding the measurement of $\mathcal A$ using non-Gaussian information of the CMB polarization rotation field. It will be valuable to study whether more realistic string networks based on physical simulations have more non-Gaussian information that can be exploited to significantly improve parameter inference.

The models we used to generate string network realizations in this paper assume that the axion string network evolves according to a scaling solution, for which the effective string length per Hubble volume $\xi_0$ is constant. Recent simulations cast doubt on this assumption, with numerical evidence of a logarithmic violation~\cite{Gorghetto_2018,Buschmann_2022}. This implies a redshift-dependent $\xi_0$, which can be accounted for in the loop-crossing model by adopting a generally redshift-dependent radius distribution $\chi(\zeta,\,z)$. The analysis of this paper could then be straightforwardly extended by replacing $\xi_0$ with a different choice of dimensionless characteristic length scale upon which $\xi_0(z)$ would depend.

By the same token, a string-wall network that collapses at a time before the present day depending on the axion mass $m_a$ can also be described by a redshift-dependent radius distribution $\chi(\zeta,\,z)$ and therefore analyzed by the technique in this paper.

\section{Conclusion}\label{sec:conclusion}

We have presented the first demonstration using mock axion string network realizations that the Peccei-Quinn-electromagnetic anomaly coefficient $\mathcal A$ can be measured from axion string-induced cosmic birefringence signals with future experiments comparable to the recently conceived CMB-HD concept. This is achieved by applying scattering transform to the anisotropic polarization rotation pattern that is estimated using quadratic estimators, and extract the non-Gaussian spatial information therein. This information has been hitherto inaccessible through the traditional power spectrum analysis, which suffers from a degeneracy between $\mathcal A$ and the effective number of string loops per Hubble volume, $\xi_0$. At the experimental capability of CMB-HD, the likelihood-based parameter inference procedure described in this paper is able to clearly distinguish high-energy-scale physics that has $\mathcal A=\sfrac{2}{3}$ from ones that have $\mathcal A=\sfrac{1}{9}$ or $\sfrac{1}{3}$, and marginally distinguish $\mathcal A=\sfrac{1}{9}$ from $\sfrac{1}{3}$. In the event that a axion string network is detected by next-generation CMB experiments such as Simons Observatory or CMB-S4, the technique explored in this work will provide crucial insight into the nature of charge quantization in new physics beyond the Standard Model.

\section*{Acknowledgments}

We thank Biwei Dai and Uros Seljak for useful discussions. We especially thank Junwu Huang for valuable comments that led to great improvement in the quality of this work, and Sihao Cheng for helping us with the use of the scattering transform method. LD acknowledges research grant support from the Alfred P. Sloan Foundation (Award Number FG-2021-16495). SF is supported by the Physics Division of Lawrence Berkeley National Laboratory.


\bibliographystyle{JHEP}
\bibliography{references}



\end{document}